\documentclass{aa}  

\usepackage{graphicx}   % Including figure files
\usepackage{amsmath}    % Advanced maths commands
\usepackage{amssymb}    % Extra maths symbols
\usepackage{multicol}        % Multi-column entries in tables
\usepackage{bm}         % Bold maths symbols, including upright Greek
\usepackage{pdflscape}  % Landscape pages
\usepackage{subfig}
\usepackage{tabularx}
\usepackage{tabularray}
\usepackage[flushleft]{threeparttable}
\usepackage{supertabular}
\usepackage{xcolor, float}
\usepackage{txfonts}
\usepackage{tipa}
\usepackage{hyperref}
\usepackage{pdflscape}
\hypersetup{
  colorlinks   = true, 
  urlcolor     = blue, 
  linkcolor    = blue, 
  citecolor    =[RGB]{255, 104, 68}
}

\usepackage{xspace}

\newcommand{\name}{{ToPP}\xspace}

\defcitealias{KJ2007}{KJ2007}
\usepackage[T1]{fontenc}
\usepackage{ae,aecompl}
\usepackage{newtxtext,newtxmath}
\begin{document} 

\title{Topology of Pulsar Profiles (ToPP)}
\subtitle{$\textsci$. Graph theory method and classification of the EPN}

\author{D.~Vohl %\orcid{0000-0003-1779-4532}
          \inst{1,2}\fnmsep\thanks{Email: d.vohl@uva.nl}
          \and
          J. van Leeuwen %\orcid{0000-0001-8503-6958} 
          \inst{1}
          \and
          Y. Maan %\orcid{0000-0002-0862-6062} 
          \inst{3, 1}
          }

   \institute{ASTRON, Netherlands Institute for Radio Astronomy, 
            Oude Hoogeveensedijk 4, 7991\,PD Dwingeloo, The Netherlands
         \and
            Anton Pannekoek Institute for Astronomy, University of Amsterdam, 
              P.O. Box 94249, 1090 GE Amsterdam, The Netherlands
         \and
            National Centre for Radio Astrophysics, Tata Institute of Fundamental Research, Pune, 411007, Maharashtra, India
             }
\authorrunning{Vohl~et~al.}

 \abstract{
Some of the most important information on a radio pulsar is derived from its average pulse profile.
Many early pulsar studies were necessarily based on only a few such profiles. 
In these studies, discrete profile components were linked to emission mechanism models for individual stars through human interpretation. 
For the population as a whole, profile morphology must reflect the geometry and overall evolution of the radio emitting regions.
The problem, however, is that this population is becoming too large for individual intensive studies of each source.
Moreover, connecting profiles from a large collection of pulsars rapidly becomes cumbersome.
In this article, we present {\name}, the first-ever unsupervised method to sort pulsars by profile-shape similarity using graph topology.
We applied \name to the publicly available European Pulsar Network profile database, providing the first organised visual overview of multi-frequency profiles representing 90 individual pulsars.
We found discrete evolutionary tracks varying from simple single-component profiles at all frequencies 
towards diverse mixtures of more complex profiles with frequency evolution.
The profile evolution is continuous, extending out to millisecond pulsars, and does not fall into sharp classes. 
We interpret the profiles as being a mixture of pulsar core-cone emission type, spin-down energetics, and the line-of-sight impact angle towards the magnetic axis.
We show how \name can systematically classify sources into the Rankin empirical profile scheme.
\name comprises one of the key unsupervised methods that will be essential to exploring upcoming pulsar census data, such as the data expected from the Square Kilometer Array.}


   \maketitle

%%%%%%%%%%%%%%%%%%%%%%%%%%%%%%%%%%%%%%%%%%%%%%%%%%

%%%%%%%%%%%%%%%%% BODY OF PAPER %%%%%%%%%%%%%%%%%%

\section{Introduction}
\label{sec:intro}

A pulsar is a rapidly rotating and highly magnetised neutron star. When there is misalignment between the spin axis and
the dipolar magnetic field axis, with angle $\alpha>0^{\circ}$, and given an adequate configuration with our
line of sight,\footnote{We refer to the angle of closest approach of the line of sight to the magnetic axis as $\beta$,
  and $\zeta=\alpha+\beta$.} a pulsar can be detected by radio telescopes as periodic short bursts of energy.
While individual pulses are highly variable for any given pulsar on the time-span of an observation,
integrating over a sufficiently large number of rotations and folding temporal observations modulo the spin period $P$ produces a pulse profile (intensity or flux density as a function of the rotation phase $\phi$) that, to the zeroth order, appears to be stable over decades of observation~\citep{Johnston+2021MNRAS.502.1253J}.
The integrated pulse profile of radio pulsars represents one of their most informative traits and is a practical tool to gain insights about key properties determined by time-stable factors such as the geometry or dominance of the strong magnetic field~\citep{LorimerKramer2004hpa..book.....L}.

Linking the pulsar rotational and energetic properties that make up its `identity' ($P$, spin-down rate $\dot{P}$, estimated characteristic age $\tau_c \approx P/(2\dot{P})$, surface magnetic field strength $B \propto \sqrt{P\dot{P}}$, or spin-down energy $\dot{E} \propto \dot{P}/P^3$) to its 
geometry and exact radio emission pattern remains difficult~\citep{Karastergiou2015aska.confE..38K}.
The known pulsar population displays pulse profiles of diverse morphology,\footnote{Throughout the article, we use the terms `shape' and `morphology' interchangeably when discussing average profiles.} where each pulsar seems to have its own multi-frequency profile signature.
If we exclude imprints on the profile morphology of propagation effects, such as pulse broadening impeded by the interstellar medium, profiles generally consist of a small number of Gaussian-like shaped components 
\citep{kramer+1994A&AS..107..515K}; profiles sometimes exhibit a slimmer shape that can be fitted with Lorentz profiles, especially at higher energies~\citep[X-rays and $\gamma$-rays; e.g.][]{Ferrigno2023A&A...677A.103F}.

Phenomenological studies have considered a form of classification based on the number of distinct components~\citep[single, double, triple, multiple/complex;][]{Backer+1976Natur.263..202B} and their polarimetric properties~\citep{Rankin1983ApJ...274..333R} with discriminants such as the location of the circular polarisation sign change, which is indicative of the crossing between the line of sight and the magnetic pole~\citep[e.g.][]{Weisberg1999ApJS..121..171W, Johnston2006MNRAS.368.1856J}.
Individual components have been studied over varying frequencies in the radio band, allowing the evolution of the beam via tracers, such as the component width, the phase separation between components, and spectral index, to be tracked.

Pulsar models have been developed to explain the source of components in relation to the magnetosphere.
Given that $P$ defines the extent of the pulsar light cylinder,
the seed plasma responsible for the pulsar emission is considered to take form on the polar cap  --  the area at the surface of the neutron star containing open field lines. The beamed radiation is emitted by the plasma streaming along the active field lines.
In the curvature radiation model~\citep{Gangadhara2004}, emission propagates tangentially to dipolar magnetic field lines.
In this picture, active field lines are expected to be circularly symmetrical about the magnetic axis, resulting in a hollow cone of emission from the last open field lines with a radius that increases with height from the star surface, and with emission height decreasing closer to the magnetic axis.

When observed at frequencies below about one gigahertz, it is common to observe an evolution of pulse width as a function of frequency: Pulsar profiles tend to be wider at lower frequencies than at higher frequencies.
Moreover, in double or multiple profiles, the outer components tend to show a greater outwards shift in phase with a decreasing frequency than those nearing the profile centroid~\citep{mitra_rankin_2002ApJ...577..322M}.
This phenomenon is generally interpreted as a given frequency corresponding to a specific beam radius and emission height, and it is colloquially known as radius-to-frequency mapping ~\citep[RFM; see e.g.][]{Cordes1978ApJ...222.1006C, Thorsett_1991ApJ...377..263T, mitra_rankin_2002ApJ...577..322M}.
This evolution is expressed using the full width at half maximum (W50) or at ten percent of maximum (W10).
Some pulsars match different phenomenological classes 
at different  frequencies, where one or more components appear or vanish with varying frequencies~\citep[see e.g.][]{Johnston+2008MNRAS.388..261J}.
These frequency-dependent characteristics have helped in determining where in the magnetosphere the radio emission arises.

To explain the symmetrical features observed in many pulsar profiles, the core and hollow cones model~\citep[][]{Rankin1983ApJ...274..333R, mitra_rankin_2002ApJ...577..322M} was proposed, comprising a core component of pencil-beam shape positioned along the magnetic axis surrounded by one or more concentric hollow cones.
Alternatively, to explain complex and asymmetrical profiles, proposed models include the patchy model~\citep{Lyne_Manchester_1988MNRAS.234..477L}, where emission occurs at random locations within the beam convolved with an annular window function, and the fan beam model~\citep{Michel1987ApJ...322..822M, Dyks2010MNRAS.401.1781D, Wang2014ApJ...789...73W}, where the emission region is constrained to elongated streams that follow the paths of a group of active field lines. Recent observations of PSRs J1906+0746 and J1136+1551~\citep[][respectively]{Desvignes2019Sci...365.1013D, Oswald2019MNRAS.489..310O} seem to favour the latter model, at least for these pulsars.

The abovementioned Rankin morphology scheme comprises five broad classes: 
core-single (${\rm S_{t}}$) pulsars dominated by a central beam emission; 
core-cone triples (T) with a clear core component sandwiched by two conal outriders; pulsars in class multiple (M) with generally five components, including a central core component and distinct inner and outer conal components;
conal-singles (${\rm S_{d}}$) with emission originating from a highly non-central trajectory across a hollow-conical emission beam; 
and conal double (D) pulsars with two prominent conal components and broad low-amplitude vestigial core emission in between that tends to become weaker at high frequency.
Furthermore, Rankin has discussed a potential quadruple (Q) class where core emission has stopped in M profiles. 
A similar evolutionary behaviour may occur in T profiles.
These classes are not fixed for a given pulsar but are rather part of a morphological continuum where a pulsar class will evolve into other classes. 
Specifically, there are three typical paths as a function both of radio frequency and orientation with respect to the line of sight: (1) ${(T?) \to \rm D \to S_d}$; (2) ${\rm S_t \to T \to D}$; and (3) ${\rm S_t \to T \to M \to (Q?)}$. 
These paths give ${\rm S_{t}}$ (core, connected to T) and ${\rm S_{d}}$ (cone, connected to D) their name.

By investigating the relation between profile complexity and three physical properties ($P$, $\dot{E}$, $\tau_c$),~\citet[][hereafter KJ2007]{KJ2007} visually classified members from an unbiased sample of 283 normal pulsar profiles ($P>20$ ms) with a high S/N as belonging to one of three categories: single, double, or complex.
The young, fast-spinning ($P<150$ ms), highly energetic ($\dot{E}>10^{35}\,{\rm erg~s^{-1}}$) pulsars from the sample did not display complex profiles, with a 60:40 ratio between single- and double-component profiles. 
Complex profiles appeared abruptly at $P \sim 150\,{\rm ms}$, $\tau_c \sim 10^5\,{\rm yr}$, and/or $\dot{E} \lesssim 10^{35}~\mathrm{erg\,s^{-1}}$. 
Beyond this transition point, the older, slower spinning, less energetic pulsars included complex profiles, with relative fractions between single-, double-, and multiple-component profiles being roughly 45:25:30.
Further simulations estimated limits on emission height range for the young and old pulsars, with the young pulsars emitting at a narrow range of high altitudes ($\sim$950--1000 km above the star surface), and the older population emitting anywhere between 20 to 1000 km, with a variable number of active regions at a given height.

\citet{Weltevrede+Johnston_2008MNRAS.391.1210W} investigated morphological differences between low and high $\dot{E}$ pulsars by counting the number of components fitted in total intensity profiles. 
Results hinted at subtle differences in profile complexity. 
However, a negligible complexity difference was identified between low and high $\dot{E}$. 
For example, pulsars J1034$-$3224 and J1745$-$3040 were classified as the most complex by this method and were indeed visually complex; while others such as J1302$-$6350 were also ranked as complex while being judged as double by eye, though with highly asymmetric components, which required fitting relatively many components.
Component asymmetry may hint at a more complex beam composition that is perhaps caused by discrete emission regions at various heights of~\citetalias{KJ2007} or originates from geometrical effects such as the impact angle $\beta$ (e.g. Figure 5 of~\citetalias{KJ2007}) and related aberration and retardation of the components~\citep{GG2003ApJ...584..418G}, or it may even have an instrumental origin (e.g. an insufficient sampling rate, leading to unresolved short-duration components).

While there has been considerable effort to determine the origin of components seen in individual pulse profiles via directed methods, we note that there has been little effort to systematically compare the shape of pulsar profiles using automated methods.
As the full Square Kilometer Array is expected to increase the population of currently known pulsars ($\sim$~3,400) by more than ten fold, a large fraction of which should already be found during Phase I~\citep{keane_2017, kramer+stapper2015aska.confE..36K}, it is timely to research and develop automated methods for knowledge extraction that require little to no human intervention.

In the following sections, we present {\name} (Topology of Pulsar Profiles),
an unsupervised method employing graph theory that can automatically compare and organise a collection of pulsar profiles based on similarities in their morphology and that enables investigation of relations between pulsar profiles and their physical properties. 

As a first case study, we applied \name
%our method
to the publicly available data from the European Pulsar Network (EPN) database. 
From that database, multi-frequency total-intensity profiles can be reliably and relatively homogeneously extracted. 
Since the start of our work~\citep[see][]{vohl_2021_7703943b, vohl_2021_7704065b}, graph theory has also been used on the 
$P$$-$$\dot{P}$ diagram \citep{2022MNRAS.515.3883G,2023MNRAS.520..599G} in isolation. 
Here we focus on the combined set of multi-frequency profiles and the pulsar spin parameters, whose combined information remains largely unexplored.
We find discrete evolutionary tracks varying from simple single-component profiles towards a diverse mixture of more complex profiles -- interpreted as a mixture of pulsar class evolution and line-of-sight impact angle towards the magnetic axis.
We investigate how to utilise the information encoded in the graph topology to predict the class of unknown pulsars and explore relations between morphology, graph locality, and physical parameters. 

In this article, we thus limited the scope of our application to morphological comparisons of total-intensity profiles over a number of frequencies. 
We plan to apply the method to other large datasets in the near future, especially those that include homogenous polarisation information~\citep[e.g.][]{2023MNRAS.520.4582P}, and to collections of fast radio bursts~\citep{2019A&ARv..27....4P, 2019ARA&A..57..417C, 2021ApJ...923....1P}.

The next sections of this article are organised as follows: 
Section \ref{sec:organizing} describes the method and the test data. 
Section \ref{sec:results} describes the experimental results. 
Finally, Section \ref{sec:discussion} discusses the analysis results, their potential physical implications, lessons learned, and future prospects.

\section{Organising pulse profiles}
\label{sec:organizing}

In this section, we describe our methodology for data acquisition,  preparation, and  analysis.

\subsection{Data acquisition from a semi-structured source}

The EPN database is a rare public collection of curated multi-frequency pulsar profiles. The database provides access to 2458 integrated profiles for 840 pulsars, taken from 77 individual publication references\footnote{See \url{http://www.epta.eu.org/epndb/about.html} for a list of EPN references. Last visit on 15 March 2021.} spanning over three decades. 
Profiles are submitted for inclusion to the database on a voluntary basis. 
Each pulsar included in the database has one or more profiles taken at specific observing frequencies originating from one or more references.
Observations can be visualised interactively in a web browser, and corresponding files can be downloaded. 

Profiles are submitted by authors in a range of data formats, which have been curated into semi-structured files of \texttt{JSON}\footnote{\url{https://www.json.org/json-en.html}, last visited 19 April 2021.} format. We collect\footnote{Using shell command \texttt{lftp -c 'mirror -c --parallel=100 http://www.epta.eu.org/epndb/json ; exit'}.} each pulsar observation appearing on the website. Each file includes a mix of metadata, and a number of curated arrays corresponding to the phase, and total intensity (stokes I). 
Additionally, some observations also provide linear and circular polarisation information (stokes $\mathrm{L}=\sqrt{\mathrm{Q}^2 + \mathrm{U}^2}$ and stokes V).
Unfortunately, the number of  reasonably reliable, calibrated polarisation profiles is small, and we do not use polarisation information in  the  current study. 
For each pulsar, we also collect\footnote{We employ the python package \texttt{psrqpy}, version 1.0.9 (\url{https://psrqpy.readthedocs.io/en/latest}, last visited 23 March 2021.)} measured and derived properties from the ATNF pulsar catalogue~\citep{Manchester+2005AJ....129.1993M}.

\subsection{Data preparation}
\label{sec:dataprep}
After collecting information for each pulsar profile from the EPN website, we link each observation to its corresponding file, and store this meta-information (pulsar name, reference, observing frequency, and file location) into tabular data\footnote{We employ the python package \texttt{Pandas}~~\citep{reback2020pandas}, version 1.6.2  (\url{https://pandas.pydata.org}, last visited 19 April 2021.)} to ease processing and analysis, and reduce the odds of introducing human errors.

As mentioned in Section \ref{sec:intro}, the pulse profile is composed of a given measured quantity (intensity or flux density) as a function of $\phi$ sampled at $n$ discrete intervals $\Delta_{\phi}=\frac{360^{\circ}}{n}$.  Within the EPN sample, profiles are sampled in phase at various rates varying between 26 to 4096 bins as provided by the various authors,\footnote{We refer the reader to individual references listed in Table \ref{tab:pulsars} for details about specific profiles' original binning procedure. Public EPN profiles are sampled at regular intervals for phase range [-0.5, 0.5].} with a majority of profiles near 512 bins.\footnote{Figure \ref{fig:epn-bins} shows the distribution of phase bins in the sample.} We re-grid profiles to a common resolution of 512 bins by applying a spline interpolation using the \texttt{zoom} function from the \texttt{ndimage} module of the \texttt{Scipy}\footnote{Version 1.10.0}~\citep{2020SciPy-NMeth} python package.

We normalise stokes I intensities for all profiles to a common range of [0, 1] to allow comparison between pulsars.
Furthermore, we apply a median correction by subtracting the profile median intensity. To compute the S/N of a profile array {$\bm X$}, we evaluate the signal as the maximum value of the stokes I profile, and the noise by splitting the 512 phase bins of the stokes I profile into chunks of 8 bins and evaluating the median of medians ($\mu$) and median of standard deviations ($\sigma$). The S/N is then computed using the following equation:
\begin{equation}
    \mathrm{S/N} = \frac{\max(\bm{X}) - \mu}{\sigma}.
\end{equation}

Multi-frequency profiles in our subset have been pre-aligned in phase for any given pulsar in the EPN, with the exception for pulsars J1803$-$2137 and J1857+0943, which we align by minimising the sum of root-square differences between their respective profiles.

To investigate the profile evolution over frequency, we divided observations into a set of discrete bins of observed frequency ranges in megahertz, hereafter frequency bin, following: $[400, 700)$, $[700, 1000)$, $[1000, 1500)$, and $[1500, 2000)$. We then created a subset of pulsars that meet the following selection criteria:
(1) Observations include at least one observation for each frequency bin; (2) each observation has a peak S/N greater than 20; (3) the profile at the lowest frequency bin does not display an obvious scatter broadening tail.

\begin{figure}
\centering
\includegraphics[width=\columnwidth]{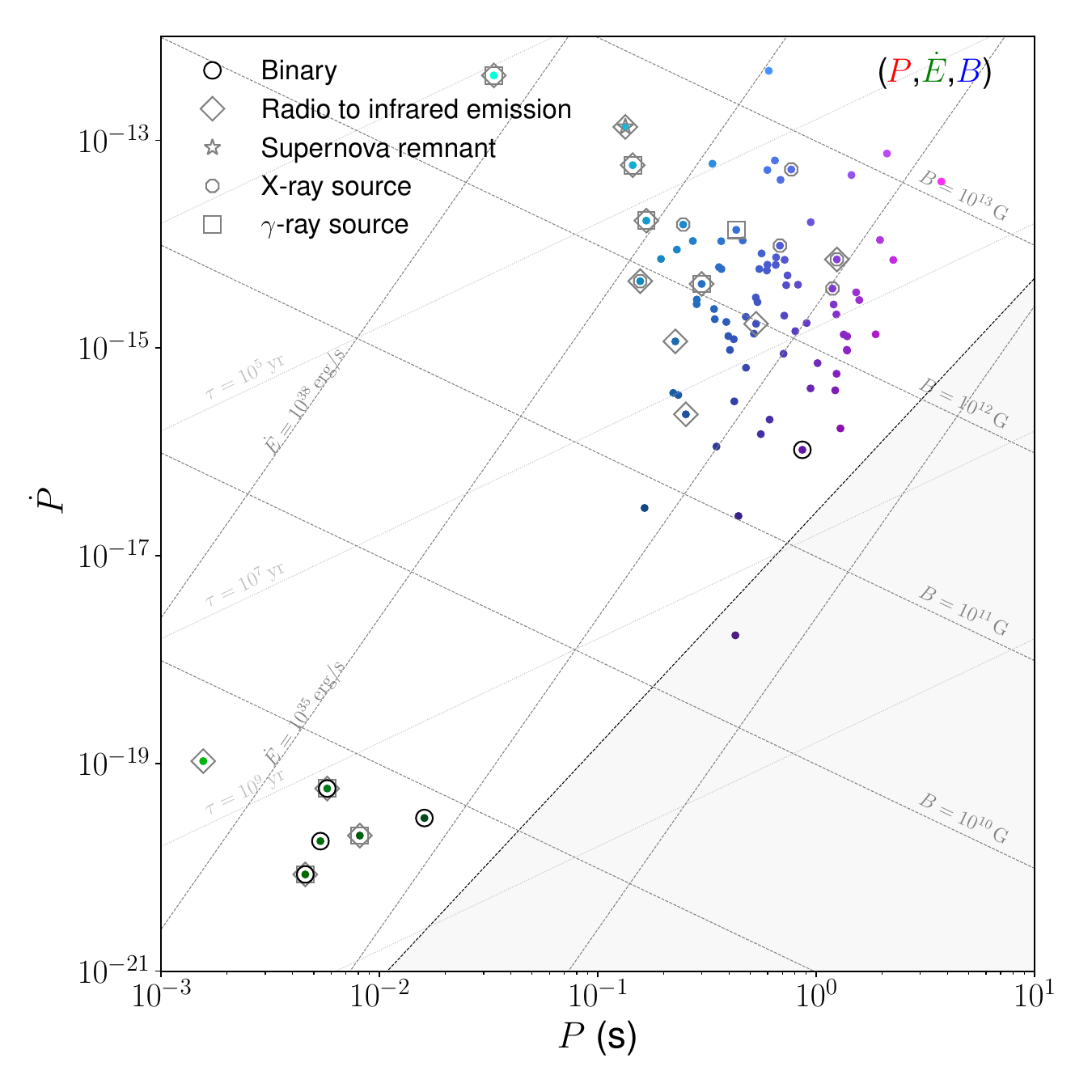}
\caption{$P-\dot{P}$ diagram for selected pulsars. Marker colours were defined by mapping the red, green, and blue (R, G, B) channels to the period, spin-down energy, and magnetic field strength ($P$, $\dot{E}$, $B$), respectively, using inverse hyperbolic sine scaling. Additionally, we highlight pulsars in binary systems, spin-powered pulsars with pulsed emission from radio to infrared or higher frequencies, association to a supernova remnant, and x-ray and $\upgamma$-ray sources.} 
\label{fig:p_pdot}
\end{figure}

For pulsars where more than one observation is available for a given frequency bin, the one with the highest S/N is selected. From the available 840 pulsars, a subset of 90 meets our set of criteria
and  remains for further analysis.
We show the period and period derivative distribution of these selected pulsars in Figure \ref{fig:p_pdot} and list them in Table \ref{tab:pulsars}.

\subsection{Graph}
\label{sec:graph}

\subsubsection{Definitions}

A graph $G=(V,E)$ is a structure composed of abstract mathematical objects forming the vertex set $V$ out of which object pairs can be connected forming the edge set $E$~\citep{busacker1965finite}.
We note the number of instances in a set $S$ -- its size -- as $|S|$.
We represent an edge $e$ connecting two vertices $u$ and $v$ as $e_{uv}=\{u,v\}$.
An edge is directed if the order by which the two vertices it connects are visited is of importance, and undirected otherwise.
The density of $G$ is defined as $D=\frac{2|E|}{|V|\cdot(|V|-1)}$.
$G$ is said to be complete if $D=1$; that is, if every vertex is directly connected by an edge to every other vertex in $V$.
A weight $w$ can be associated with an edge to consider a measure of distance separating its two associated vertices; in such a case, given each edge in $E$ has an associated $w$, we also obtain the weight set $W$.
We denote the exclusion of a vertex $u$ from the set $V$ as $V\setminus u$.
The degree $\deg(v)$ of a vertex $v$ in an unweighted graph corresponds to the number of edges incident to $v$, while the degree in a weighted graph can be represented as the sum of edge weights incident to $v$.
Following the unweighted degree definition, we refer to a vertex $u$ of $deg(u)=1$ as a leaf.

A tree is a special kind of connected graph for which there exists a series of edges connecting any two vertex -- a path\footnote{In an undirected graph, a series of edges is usually referred to as a chain; however, we chose to the term path in this work for language simplicity.} -- where following consecutive edges in the set (e.g. $[e_{ab}, e_{bc}, e_{cd}, ...]$) does not cycle (e.g. cannot be of the form $[e_{ab}, e_{bc}, e_{ca}]$).
The distance between two vertices corresponds to the number of edges in the path connecting one to the other.
For a weighted graph, the distance can include the sum of the visited edges' weights.
The shortest path between two vertices is the set of edges in a graph that minimises their distance.
The measure of centrality of a vertex in a graph carries information about the importance of said vertex in the graph and permits identification of the most significant vertices.
In particular, we defined the closeness centrality measure~\citep[][]{freeman1978centrality} as the average of the shortest path length from a given vertex to every other vertices in $V$.
We refer to the vertex with maximal closeness centrality measure as the root.
The contraction $G/e$ of an edge $e_{uv}$ is the operation $E \setminus e$ that removes the edge connecting $u$ and $v$, merging the two vertices into a new vertex $uv$ in $V$ that is connected to all original neighbours of both $u$ and $v$.
If $u$ and $v$ are not directly connected, but rather are connected via a path, $G/e$ removes all edges forming the path, and merges the vertices it spans.

\subsubsection{Application}
\label{subsec:application}

To organise pulsars by similarity of their multi-frequency profiles, we construct a complete undirected weighted graph $G=\{V,E,W\}$ using our pulsar population, where $V$ is the vertex set of size $|V|$, $E$ is the set of $|E|=\big(\begin{smallmatrix}
  |V|\\
  2
\end{smallmatrix}\big)=\frac{|V|\cdot(|V|-1)}{2}$ edges, and $W$ is the set of $|W|=|E|$ edge weights connecting every edge in $E$ to a unique weight in $W$. We represent each pulsar from our population as a vertex\footnote{|V| = 90.}. We represent the comparison between any two pulsars $u,v \in V$ as an edge $e$ with an associated weight $w$ corresponding to their degree of similarity, forming the set $e_{uvw}=\{u, v, w\}$.

From $G$, we computed the minimum spanning tree (MST) -- a subset of edges covering all vertices, without any cycle, that globally minimises the sum of the weights -- by applying Kruskal's algorithm~\citep{kruskal1956shortest}, reducing the set of edges to size $|V|-1$.
The shape of the MST carries information regarding the metric used to define the weights, where an elongated shape shows sign of structured variations in the data, while a rather circular graph would hint towards noise \citep[in relation to the measured quantity;][]{Baron+2021ApJ...916...91B}.
To quantify the MST shape, we evaluate its minor and major axes, corresponding to the estimated half width $\hat{h}$ and the longest path length $l$ respectively.

Given the nature of a tree, the vertex in $V$ with the smallest closeness centrality measure is the least connected vertex $v_{0}$ (a leaf), which is expected to be located at one end of the longest path.
We defined the distance between any vertex $q \in V\setminus v_{0}$ and $v_{0}$ as its level $L_{v_{0}}(q)>0$, where $L_{v_{0}}(v_{0})=0$.
We defined the MST root as $v_{r}$.
When plotting the MST, we defined the root level $L_{v_{r}}(v_{r})=0$ and the level of any other vertex $q$ relative to $v_{r}$.

We computed $\hat{h}$ as the average degree for all levels $L_{v_{0}}(x)~\forall x \in V$.
The longest path along edges forms an ordered sequence of pulsar profiles.
We found this longest sequence by applying two consecutive breadth-first searches~\citep{dijkstra1959note}. First, we started from any node of the graph, and we applied a breadth-first search, which leads to a leaf $u$ -- which is one end of the longest path. Then, starting from this leaf, we proceeded with a second iteration of the breadth-first search, traversing the MST and leading to another leaf $k$.
This ordered set of edge-connected vertices $\{u,v,...,k\}$ of size $l$ covers the longest path.

We statistically summarised the MST's topology by evaluating the MST elongation $\eta$ as the ratio $l/(2\hat{h})$, which we normalised by the number of vertices in the graph, $\eta'={\small \frac{\eta}{|V|}}$.
If $\eta'$ nears unity, it is indicative that the evaluated similarity measure uncovers the presence of a significant trend in the data~\citep{Baron+2021ApJ...916...91B}.
In a case where data are categorical rather than continuous with respect to a given metric, or in the presence of multiple distinct trends, unity is not expected.

\subsection{Measuring similarity between pulsars}
\label{sec:similarity}

There are many ways in which two profiles can be compared, and these different ways can lead to very different solutions.
Several distance measures to compare time series have been discussed in the literature. \citet{esling2012time} categorise distance measures for time series based on four types: shape, feature, edit, and structure.
Shape-based distances compare the overall shape of a time series on its individual values directly.
Feature-based distances are applied to obtain a dimensional reduction and a noise reduction.
Edit-based distances compares two time series based on the minimum number of operations that are required to transform one series into the other series.
Structure-based distances are obtained by comparing higher level structures obtained by modelling or compressing the time series.
For example, dimensionality reduction techniques like principal component analysis (PCA) have been utilise for time series  classification~\citep{gianniotis2017linear}.
In this work, we will not proceed with a comparative study of distance measures, reserving such an exercise for future work.
Rather, we focus our attention on a specific shape-based measure as it naturally maps to the heuristic that individual observed components relate to discrete emission regions and geometry (\S\ref{sec:intro}). Furthermore, the shape-based measure we employ in this work is a demonstrated robust time series classification method~\citep{bagnall2017great}.

Shape-based measures comprise two sub-categories: lock-step measures and elastic measures.
Lock-step measures directly compare two time series following a one-to-one relation.
Examples include the Minkowski distance ($L_p$-norm of the difference between two vectors of equal length, and where the Euclidean distance corresponds to $p=2$, also known as $L_2$-norm) and the Pearson correlation distance.
Lock-step distance measures are sensitive to scale and time shifts.
Elastic measures have been developed to overcome these problems, and generally provide better classification accuracy compared to lock-step measures~\citep{gorecki2018classification}.
An elastic measure well discussed in the literature is the Dynamic Time Warping (DTW) algorithm~\citep{BellmanKalaba1104847}.
Whereas the $L_2$-norm compares sample $x_i$ with sample $y_i$ for every $i$, DTW also considers neighbouring samples to allow for shifts and stretches~\citep{Niennattrakul+4197360} -- which in our context allows us to abstract the effect of RFM and compare profiles' shapes invariant of their widths.
It is a similarity measure that approximately models the time-axis fluctuation (or phase-axis in our context) with a nonlinear warping function.

\begin{figure*}[h!]
\begin{tabular}{cc}
\includegraphics[trim=0 0 0 0, clip, width=8.5cm]{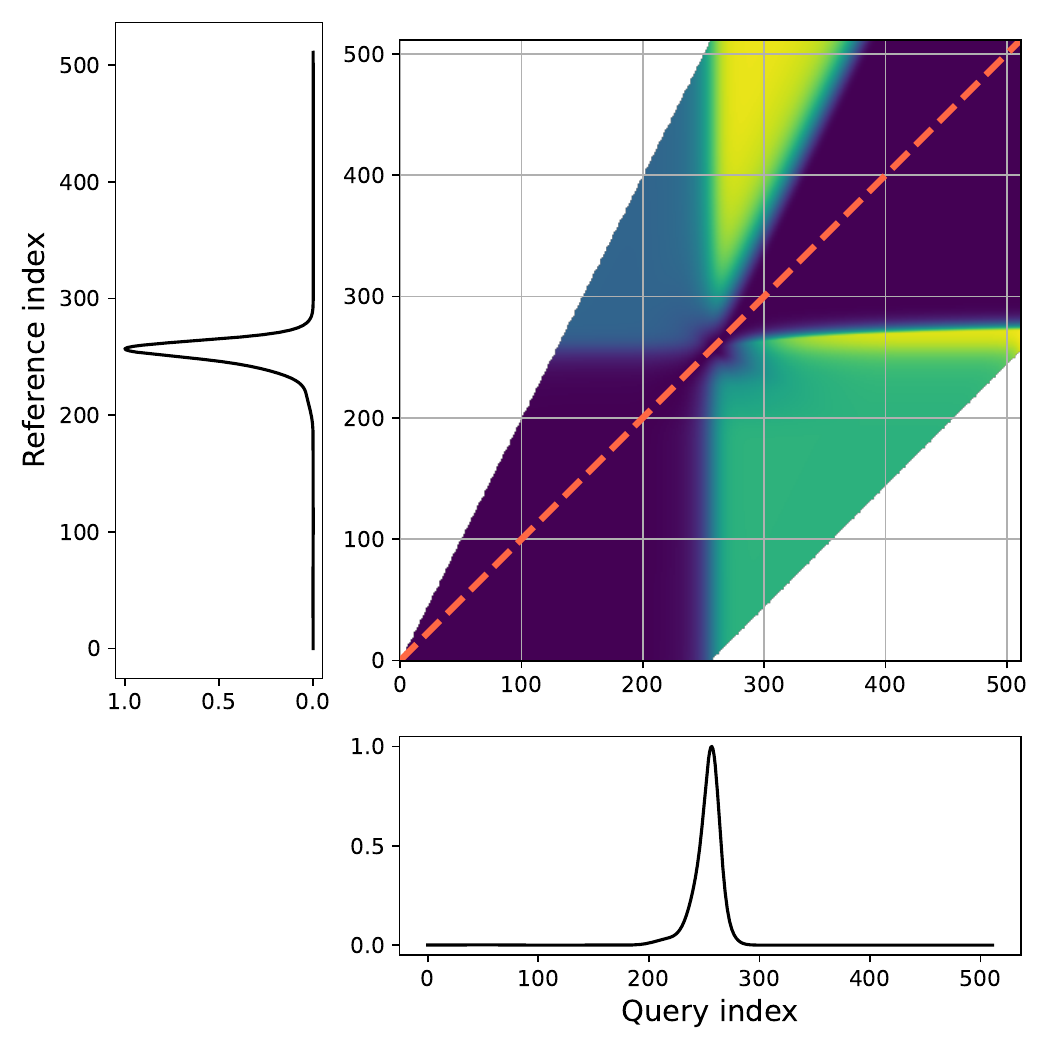} &
\includegraphics[trim=0 0 0 0, clip, width=8.5cm]{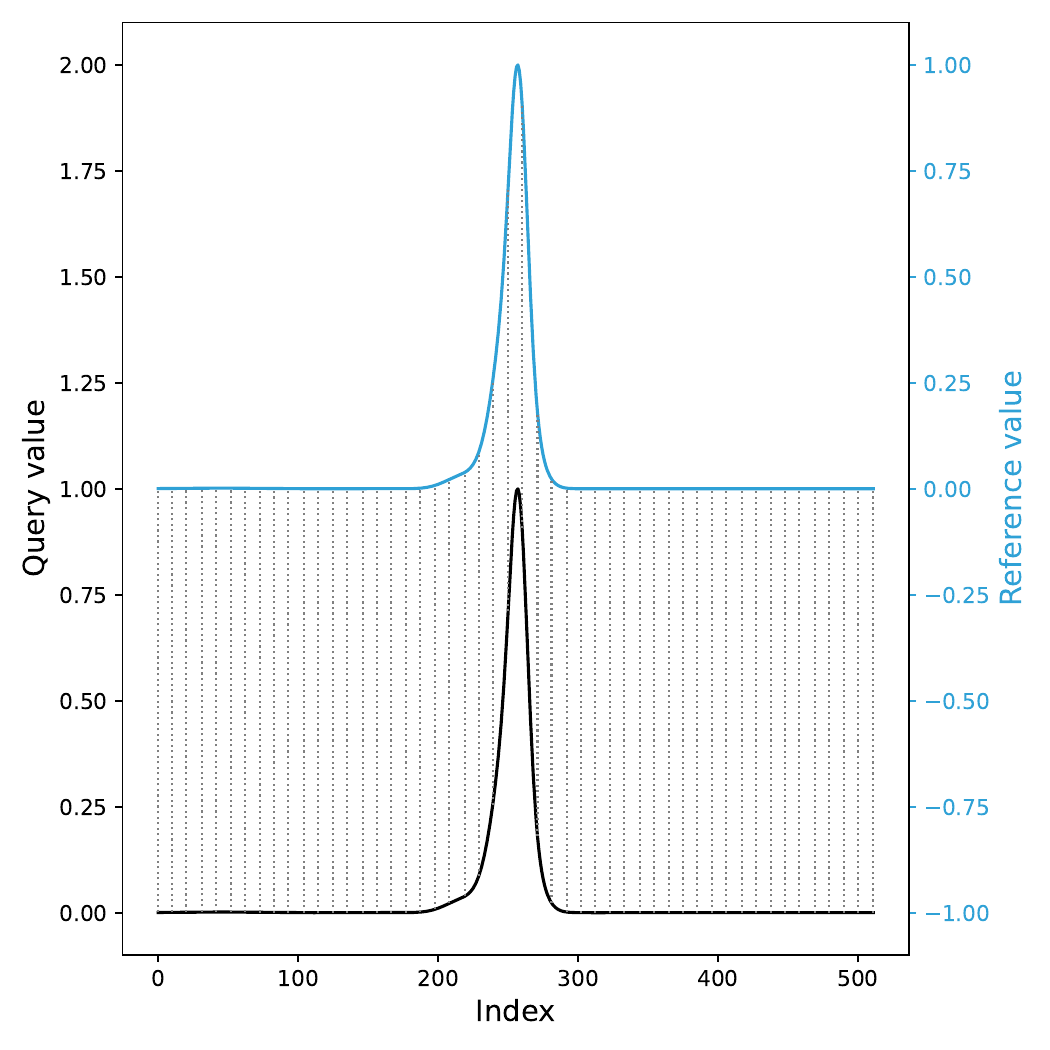} \\
\includegraphics[trim=0 0 0 0, clip, width=8.5cm]{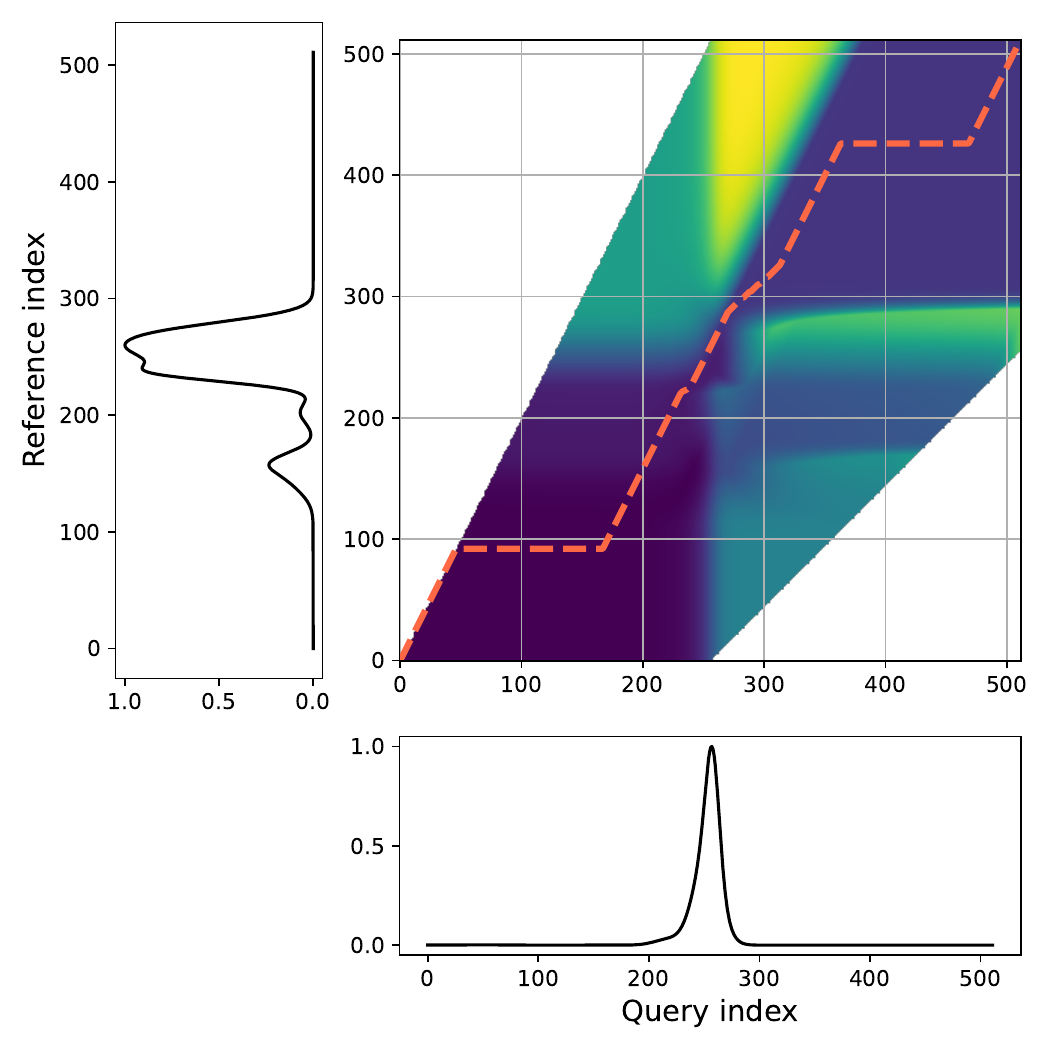} &
\includegraphics[trim=0 0 0 0, clip, width=8.5cm]{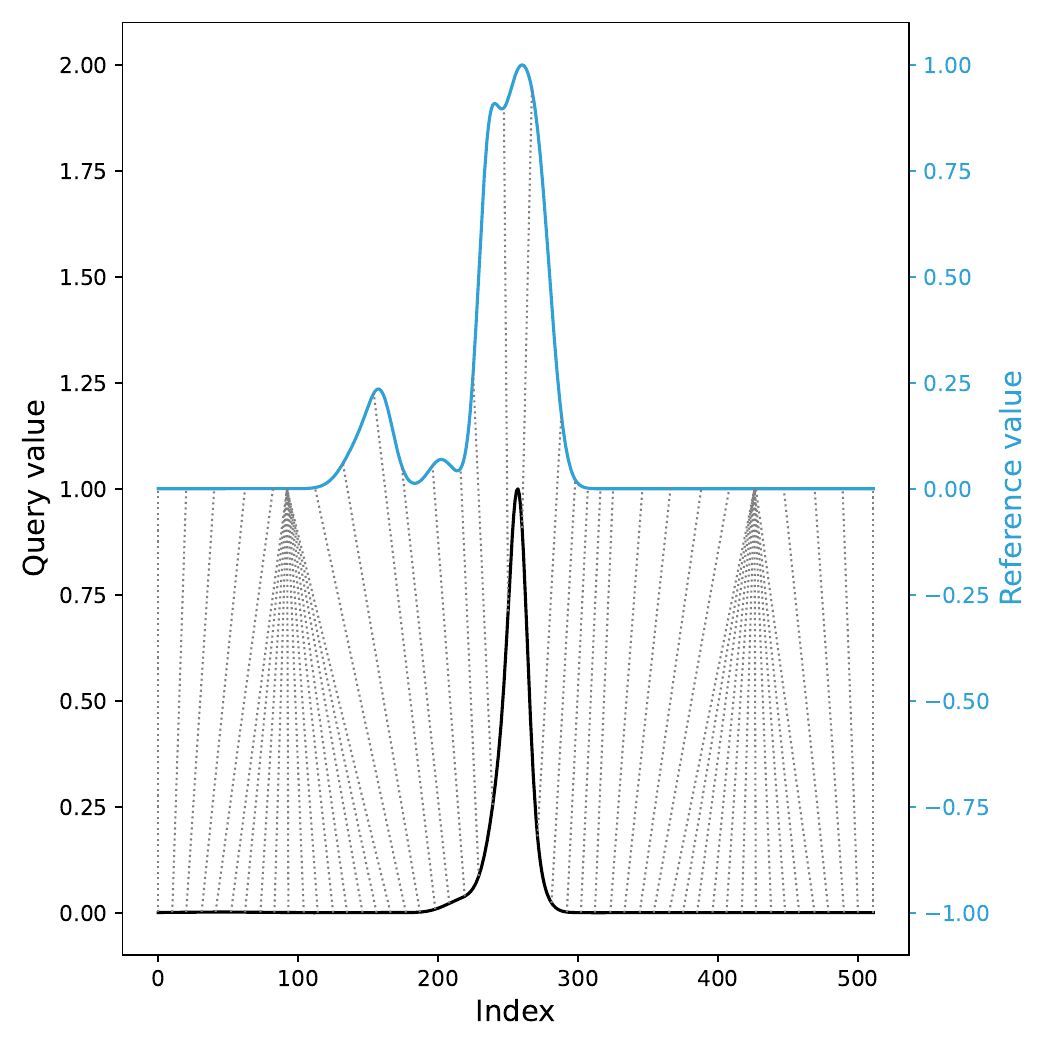} 
\end{tabular}
\caption{
Depiction of distance measurement by the dynamic time warp algorithm. 
In the left panels, pulse profiles are shown as query and reference profiles corresponding to the x and y axes of the local cost matrix where the `warping path' is computed. In the right panels, the same two profiles now show a subset of warping tracks highlighting matched samples along the warping path.
The top row shows a comparison of a profile from PSR B0950+08 (J0953+0755) with itself leading a path of least resistance (following the heat map `cost valley') corresponding to a perfect diagonal in the local cost matrix. The bottom row shows a comparison of PSR B0950+08 (J0953+0755) as a query with J1803–2137 (B1800–21) as a reference.
The path of least resistance steps off the perfect diagonal, leading to a higher distance measure.
The heat map also highlights the band of size $K=256$ used to constrain our search. 
}
\label{fig:dtw_explained}%
\end{figure*}

We briefly describe the algorithm.
DTW is a pattern matching algorithm part of the `dynamic programming' family of algorithms. The algorithm first builds an $(n \times m)$ local cost matrix (LCM), with element $(i, j)$ being the $L_2$-norm between $x_i$ and $y_j$.
The second step is to construct a warping path $W = w_1, w_2, ..., w_K$, where $max(n, m) \leq K \leq m+n-1$.
This path traverses the LCM under three constraints.
The boundary constraint enforces that the path must begin and end on the diagonal corners of the LCM: $w_1 = (1,1)$ and $w_K = (n,m)$.
The continuity constraint imposes that steps in the path are taken only among adjacent elements in the matrix, including diagonal adjacent elements.
Finally, the monotonicity constraint assures that subsequent steps in the path must be monotonically spaced in time.
The resulting total distance for path $W$ is the sum of individual LCM elements that the path traverses.
For an overview the DTW algorithm and its many variations and constraints, we refer the reader to~\citet{giorgino2009computing}, which also describe the \texttt{dtw} package used in this work, as well as details regarding constraints (namely the step pattern, window type, and window size) used to compute the warping path described below.
Figure \ref{fig:dtw_explained} exemplifies how distance is computed by DTW, with the warping curve on the LCM (left panel) and warping tracks (right panel) computed from the query and reference profiles, exemplified by PSR B0950+08 (J0953+0755) and J1735-0724 (B1732-07) respectively.

Given that a few profiles in the set present an interpulse (a secondary pulse separated by about $180^{\circ}$ from the main pulse) located near $\phi\sim-0.5$, the interpulse shape would be compared to the main pulse of another pulsar and confuse our results. 
To mitigate this problem, we applied an asymmetric step pattern and constrained the path search to a band of size $K=256$ around the main diagonal~\citep[as first described by][]{sakoe1978dynamic}, which corresponds to half a rotation. 
While the symmetric step pattern returns the average distance for both profiles, the asymmetric case returns the distance separating a query profile with respect to a reference profile. 
We therefore computed the asymmetric distance by varying the query and reference with either pulsars in a pair and selected the maximal distance (upper bound) of the two to avoid a reduced distance caused by partial matching.
We did not normalise the distance, as all profiles have the same length $n$.

During our early experiment phase, we found that for some lower S/N profiles, when DTW is applied directly to pre-processed profiles (as described in Section \ref{sec:dataprep}), it is sensitive to intensity fluctuations (e.g. noisy profiles being considered similar).
Given this observation and the fact that the EPN profiles do not include the original time series used to compute the profiles, we fit each profile using a variational Bayesian estimation of a Gaussian mixture.\footnote{We computed the mixture model with the {\tt BayesianGaussianMixture} function from the {\tt scikit-learn} python package~\citep[][version 1.2.0]{scikit-learn}} 
Using the Dirichlet process as a weight concentration prior and starting with 30 mixture components and a weight
concentration prior\footnote{We varied the number of mixture components between one and 40 to compare to the original 30 and the weight concentration prior between $10^1$ to $10^{10}$ and found little variation in the resulting models.} of $10^4$, we computed an automatic selection of active components, where the number of components to be fitted is itself modelled as a probability distribution~\citep{Ay2020MNRAS.493..713A}.
Finally, we computed the DTW distance between models.
We included a symmetry constraint in the distance measure to consider profiles with similar features but flipped in phase (e.g. bright leading component versus bright following component).
We considered this symmetry freedom well justified by the fact that the order in which  the components appear is determined by the pulsar spin direction on the sky as seen from Earth, which is independent of the emission mechanism.
We show both the pre-processed data and models in the following sections.

\section{Experimental results}
\label{sec:results}

We evaluate similarity between pulsars in our subset population by organising pulsars by pulse shape similarity. 
As mentioned in Section \ref{sec:intro}, profiles can evolve over frequency.  
To take profile morphology frequency (non-)evolution into account when comparing pulsars, we construct an MST  by averaging $w$ over multiple frequency bins.
We begin by discussing the longest sequence, expected to visualise the main evolutionary trend of the pulsar profile population.
We then investigate the full MST.

\subsection{Longest sequence}
\label{subsec:multifreq_longest}

\begin{figure*}[h!]
\centering
\includegraphics[trim=60 0 0 0, clip, width=17cm]{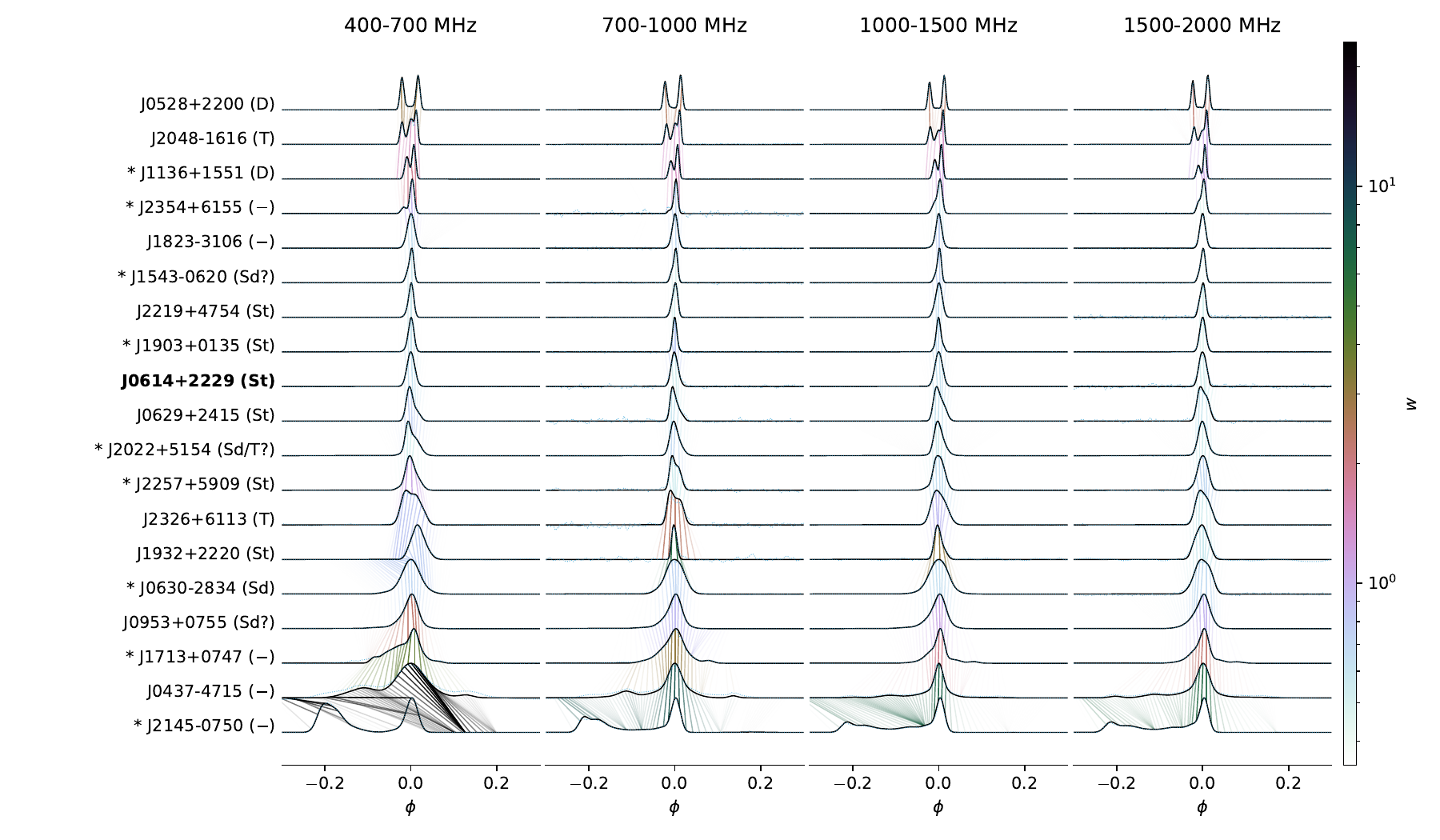}
\caption{Longest sequence optimising the MST on averaging distance from all frequency bins. The sequence includes 21\% of the original set, shown in Figure \ref{fig:tree-4freqs-w10}. 
The symmetry constraint (\S\ref{sec:similarity}) implies that neighbouring profiles in the sequence can be considered similar when inverted in phase. All sources that are mirrored with respect to the root pulsar J0614+2229 are marked with an asterisk before their name,  
and we display inverted profiles at all frequencies. 
We note that the apparent shifts in the 400$-$700 MHz bin
between the J0437$-$4715 and J2145$-$0750 profile are not, by themselves, a cause for concern, as the profiles displayed here have no underlying intrinsic physical alignment (in contrast to, e.g., a multi-frequency plot). Regardless, the black colouring of the warp between these two profiles indicates a relatively large distance.}%
\label{fig:sequence_multifreq}%
\end{figure*}

Figure \ref{fig:sequence_multifreq} shows the longest sequence within the MST, with frequency bins ordered from low frequencies (higher altitude within the magnetosphere) to high frequencies (closer to the neutron star's surface).
Each row correspond to a single pulsar with its name on the left and  morphological class, if available in \citet{Rankin1990ApJ...352..247R} or \citet{2011ApJ...727...92M}, as detailed in \S\ref{sec:classification}.
Next shown is the model profile (black curve) and the pre-processed original profile data (blue dotted curve).
The pulsar located at the root of the MST (J0614+2229) is marked in bold.
The two branches forming the sequence display length asymmetry, with 8 neighbours on one side of the root, and 10 neighbours on the other.
Profiles vary from single profiles with little frequency evolution around the root to more complex profiles at each end of the sequence.
Some pulsars show constant morphology over all frequency bins, and others show components appear, fade, or blend in with other components from one frequency bin to the next.

The vertex distribution within the tree realisation directly relates to the MST being the result of globally minimising $\Sigma w$. 
Consecutive profiles $x$ and $y$ in the sequence at a same frequency bin are connected through discrete steps of their warping path marked as vertical lines connecting point $x_i$ and point $y_j$, with transparency level varying with profile's intensity for visibility, and coloured by their atomic distance, where the averaged distance $w$ over the four bins is used to build the MST.

The longest sequence\footnote{The evolution of spin parameters $P$, $\dot{E}$ and $B$ as a function of the sequence order is presented in Figure \ref{fig:sequence-versus-spinparams}.} comprises 21\% of the pulsars in the set, 
with a normalised elongation $\eta'=0.02$ obtained from $l=11.4$ and $\hat{h}=4.7$.
Given multiple branches in the MST, the longest sequence has to be composed of the two longest branches, disjoint at the most central vertex $v_r$.
The sequence indeed shows a shape evolution distributed between two leaves of the tree, namely J2145$-$0750 and J0528+2200, located at both ends of the sequence.
The root has a single component and is classified as ${\rm S_t}$ in the literature.

Given our distance measure, the fact that most profiles are not part of the main sequence indicates a simple fact about the data: ordering the data distribution cannot simply evolve linearly, else $\eta' \to 1$.
As $\eta'$ is low, the tree must have multiple branches, with each branch gradually evolving from single profile (near the root) towards more complex morphology types near the leaves.
The assumption is that each branch must correspond to a morphology evolution track that differs from that found in other branches. 
We investigate these variations in Section \ref{subsec:multifreq_mst}.

\begin{figure*}
\includegraphics[trim=0 0 0 0, clip, width=18.5cm]{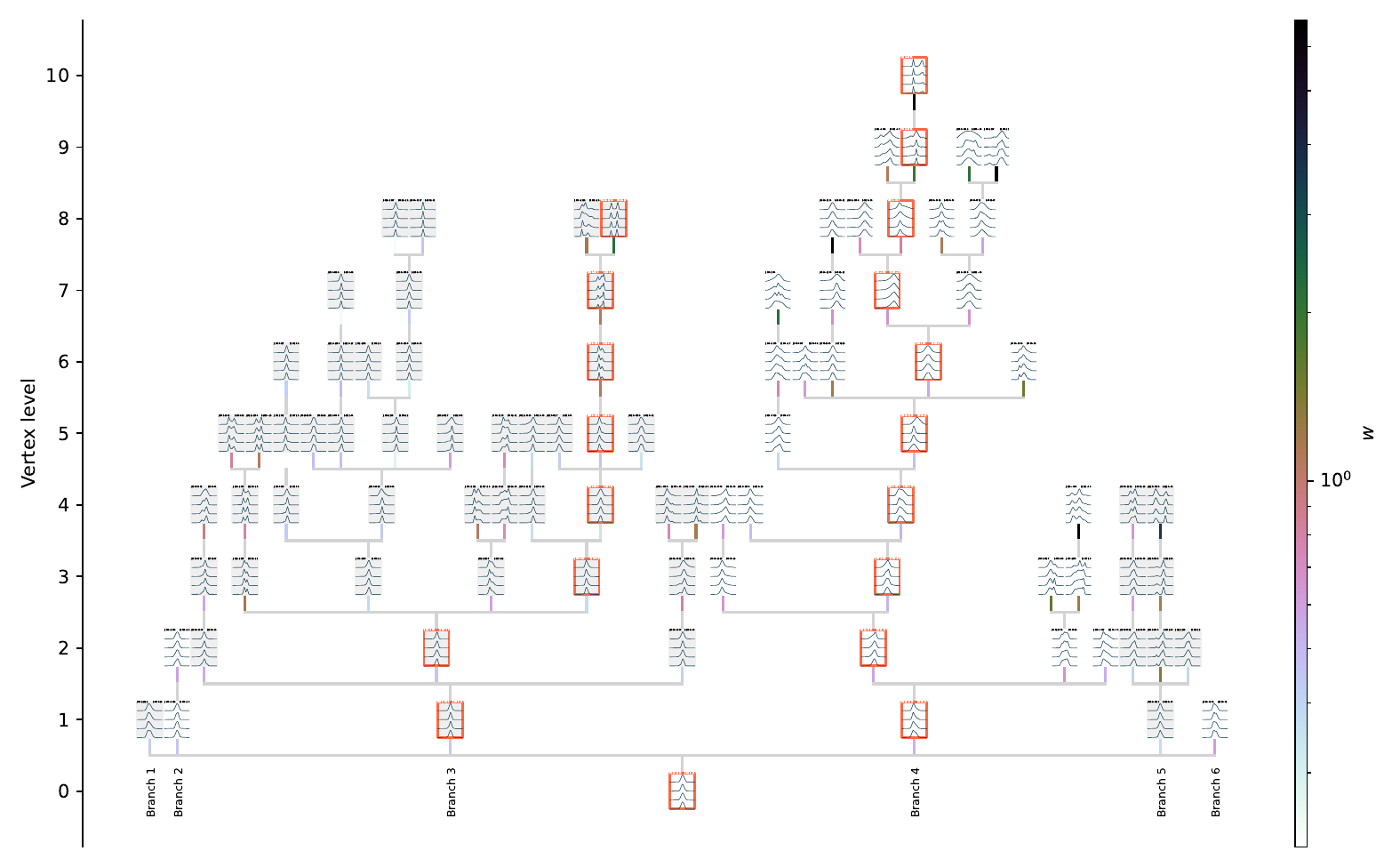} \\
\caption{Minimum spanning tree obtained using profiles by averaging distance at all four frequency bins. 
Each box shows the profiles of a given pulsar with the rotational phase showing the ten percent width (W10) plus an extra
0.05 on each side. 
The lowest frequency is at the top, and the highest is at the bottom of each panel.
The root (most central vertex) can be seen at level 0, and a child is located at level $L$(parent) + 1.
The weight $w$ (distance obtained by computing the DTW) between a child and its parent is shown as the colour of the line below each child.
Pulsars forming the longest sequence (Figure \ref{fig:sequence_multifreq}) are highlighted with orange borders. 
The digital version of this figure can be zoomed and panned to inspect individual details. 
Pulsars can be searched by name, which combined with an appropriate zoom, can ease localisation of them in the MST.
}
\label{fig:tree-4freqs-w10}%
\end{figure*}

\subsection{MST}
\label{subsec:multifreq_mst}

\begin{figure*}[h!]
\begin{tabular}{c}
\includegraphics[trim=0 0 0 0, clip, width=17cm]{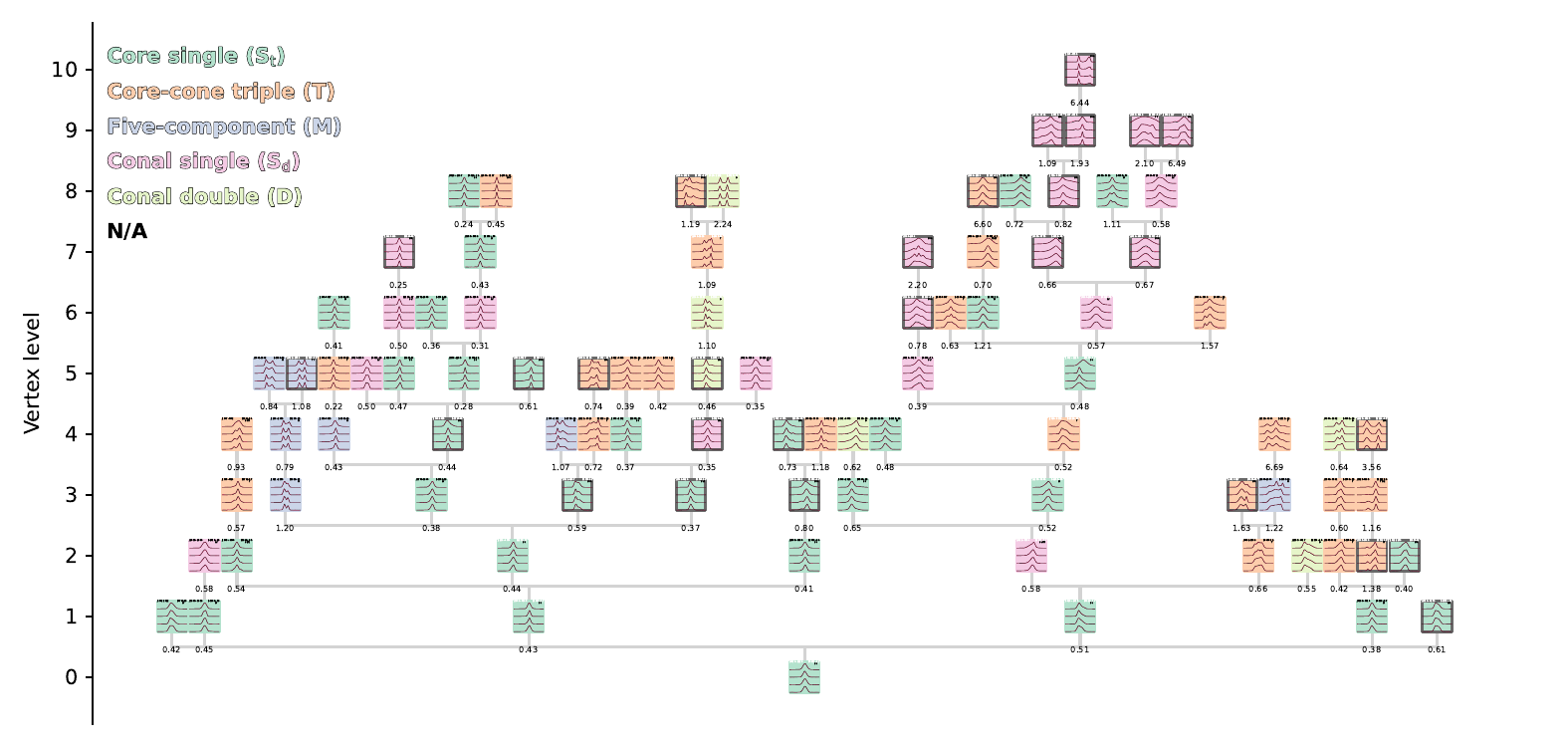} \\
\includegraphics[trim=0 0 0 0, clip, width=17cm]{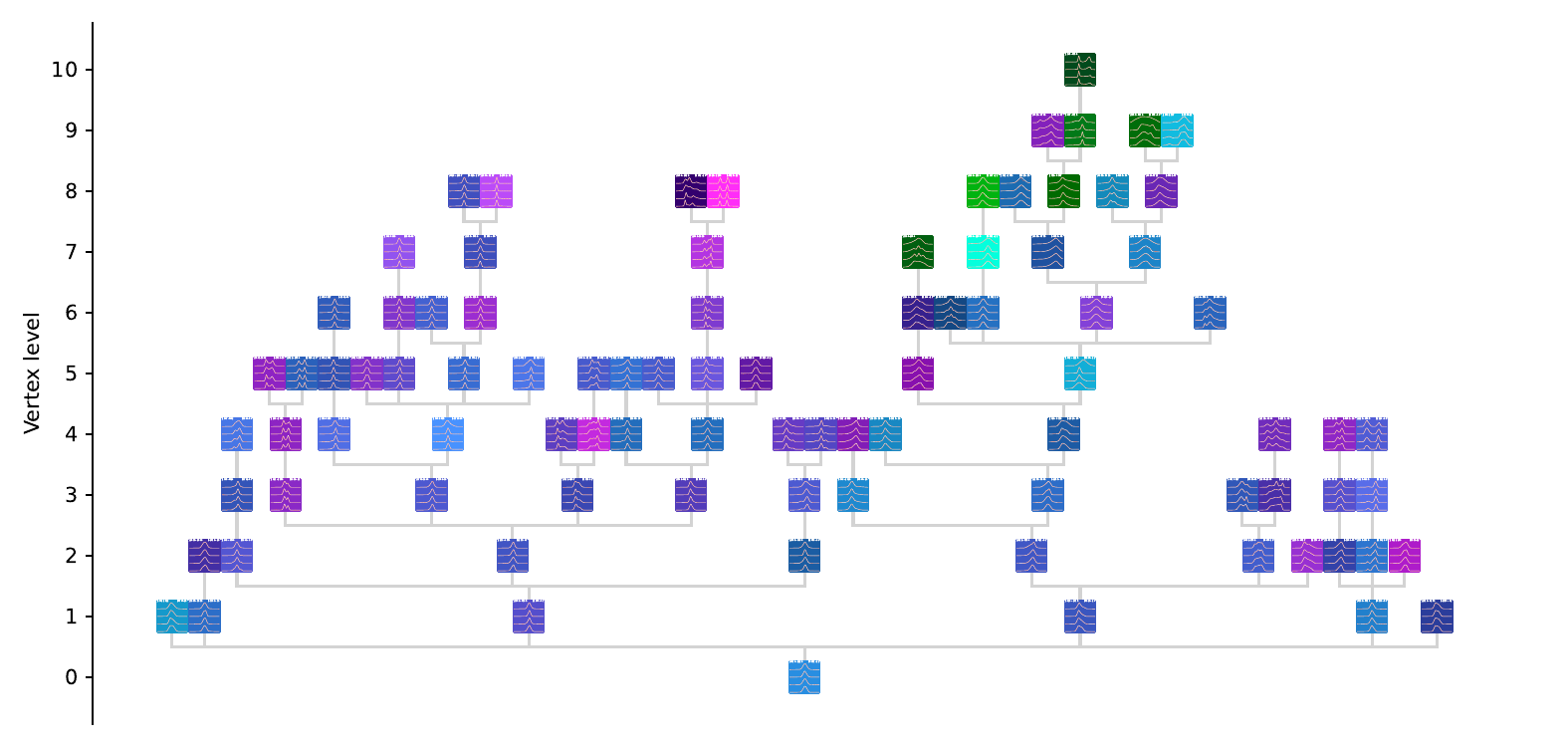} 
\end{tabular}
\caption{
Same MST as in Figure \ref{fig:tree-4freqs-w10}.
Upper panel: Rankin classification scheme. (See \S\ref{sec:classification} for details.) 
Grey borders correspond to pulsars without a clear prior classification in the reference papers; classified here using
the method described in \S\ref{sec:classification}.
To simplify classification interpretation, instead of using a colour bar to display DTW distance between neighbouring vertices, we display $w$ directly below each branch.
Lower panel: Relation between pulsar positions in the tree and in $P-\dot{P}$ space. The box colour results from mapping $P$,
$\dot{E}$, and $B$ (Figure \ref{fig:p_pdot}) to the red, green, and blue (RGB) channels, respectively.
}
\label{fig:tree-4freqs-w10-colour}%
\end{figure*}

Figure \ref{fig:tree-4freqs-w10} shows the full MST, displaying the global distribution of multi-frequency profiles, with the lowest frequency bin at the top and highest frequency bin at the bottom of each panel.
Here, all 90 pulsars are organised into branches starting from the root $v_{r}$ (as described in Section \ref{sec:graph}) at level 0.
Each vertex is represented as a box that includes the pulse profile, showing a phase range covering W10, plus an extra
0.05 on each side\footnote{Refer to Appendix \ref{app:other-msts} MST displaying full rotational phase ($\phi \in [-0.5, 0.5]$; Figure \ref{fig:tree-full}).} to highlight structure in the main pulse, with the model (solid line) and the pre-processed data (dotted line).
Each child is located at level $L_r$(parent)+1, with the DTW distance between it and its parent indicated as the colour of the connecting segment beneath it (colour bar).
Pulsars that are member of the longest sequence, as presented in Figure \ref{fig:sequence_multifreq}, are marked with a border.
We annotate each branch starting at level 1 with a numerical value (1 to 6), noting branches 1 and 6 are actually leaves.

Figure \ref{fig:tree-4freqs-w10-colour} presents two additional versions of the MST. 
In the upper panel, boxes colours are set based on the Rankin scheme with classes taken from \citet{Rankin1990ApJ...352..247R} and \citet{2011ApJ...727...92M}, respectively. 
Boxes without a classification in either reference are then classified given locality (more details in \S\ref{sec:classification}). 
The lower panel presents boxes coloured via a mapping of $P$, $\dot{E}$, and $B$ to the red, green, and blue (RGB) channels respectively. 
Hence, the longer the period, the redder a point is. 
Similarly, the higher $\dot{E}$ is, the greener; and finally the higher the magnetic field strength, the bluer.  
This mapping provides a visual cue that allows for comparison of the boxes in the MST with Figure \ref{fig:p_pdot}. 
For example, boxes in shades of pink and purple are slow pulsars with low $\dot{E}$; the pinker the stronger $B$. 
Similarly, light blue and cyan indicates slightly shorter periods and high $\dot{E}$; and so on. 
Finally, recycled pulsars are all in shades of green, given their short periods and low $B$. 

We can see from the MST structure that profiles over a range of frequencies comprise more complexity than in the case of considering a single frequency bin. 
In addition of profile morphology variation on a per frequency basis, variations of morphology (or not) across frequency bins for individual pulsar and across pulsars become discriminant. 
As we discussed in Section \ref{subsec:multifreq_longest}, each branch forms a specific morphology evolution track, starting from the simple profile $v_r$ from which various morphological types evolve.
The general trend seems to be linked to the degree of symmetry in the single 'seed' profile (at vertex level 1) that then morphs into other profile kinds.
 
The lower panel of Figure \ref{fig:tree-4freqs-w10-colour} provides hints that the relation between profile morphology and spin parameters is not purely random. 
Indeed, pulsars with similar shades (and therefore similar properties) tend to cluster, with a number of sub-branches showing parameters evolution. 
For example, this is the case for the sequence between J1823--3106\footnote{$P\sim0.284$\,s, $\log\dot{E}\sim33.7\,{\rm erg\,s^{-1}}$, $B\sim12.0$\,G}, J2354+6155\footnote{$P\sim0.945$\,s, $\log\dot{E}\sim32.9\,{\rm erg\,s^{-1}}$, $B\sim12.6$\,G}, J1136+1551\footnote{$P\sim1.188$\,s, $\log\dot{E}\sim31.9\,{\rm erg\,s^{-1}}$, $B\sim12.3$\,G}, J2048-1616\footnote{$P\sim1.962$\,s, $\log\dot{E}\sim31.8\,{\rm erg\,s^{-1}}$, $B\sim12.7$\,G}, and J0528+2200\footnote{$P\sim3.746$\,s, $\log\dot{E}\sim31.5\,{\rm erg\,s^{-1}}$, $B\sim13.1$\,G}, evolving from blue to purple to pink with monotonically increasing periods, monotonically decreasing spin-down energies, and generally increasing magnetic field strengths. 
Similarly, all recycled pulsars are found near or at the leaves in branch 4.
 
The two leaves at level 1 are occupied by PSRs~J0742$-$2822  and J2313+4253, respectively, classified in the literature as ${\rm S_t}$ and $S_d?$. 
Both stars display profile frequency evolution with a blended trailing component with increasing amplitude towards higher frequencies.  
PSR~J0742$-$2822 may therefore be a ${\rm S_t}$/T given the circular polarisation sign shift at around 600\,MHz, and J2313+4253 a conal single given its lack of sign shift in circular polarisation~\citep[e.g.][]{gl98}. 
This represents a case where polarisation would be crucial to properly distinguish profile class, as the shape and frequency evolution led to classifying J2313+4253 as $S_t$ -- supporting our desire to include this in future work (see Sect.~\ref{sec:discussion}).
Out of the four remaining branches, branches 3 and 4 comprise the bulk of the pulsars in the set; branch 5 being the third in importance, and branch 2 having a single child. 

\paragraph{Branch 2} (two vertices) is made of PSRs J1935+1616 and J2018+2839. 
These two apparently similar profiles according to DTW are also classified as core and conal singles respectively. 
As our dataset did not allow us to dependably utilise polarisation information, our algorithm is sometimes not yet capable of  differentiating between these two based on shape alone. 
While J1935+1616 has a strong sign shift in circular polarisation over a wide range of frequencies~\citep[][though displaying a second blended leading component in the 700--1000\,MHz bin]{gl98}\footnote{See, e.g., \href{https://psrweb.jb.man.ac.uk/epndb/\#gl98/J1935+1616/gl98_1408.epn}{profile at 1408\,MHz.}, last visited on 2 November 2023.}, only the circular polarisation profile of J2018+2839 at 408\,MHz\footnote{\href{https://psrweb.jb.man.ac.uk/epndb/\#gl98/J2018+2839/gl98_408.epn}{J2018+2839 profile at 408\,MHz.}} shows a sign shift below the trailing core peak. 
However, this is not seen at the three highest frequency bins of our sample, which rather show an antipodal profile (bump near pulse centroid) -- hence a ${\rm S_d}$ classification. 

\begin{figure*}[t!]
\begin{tblr}{
  colspec = {X[c]X[c]},
  stretch = 0,
  rowsep = 0pt,
  colsep=1pt
}
\includegraphics[trim=0 0 0 0, clip, width=\columnwidth]{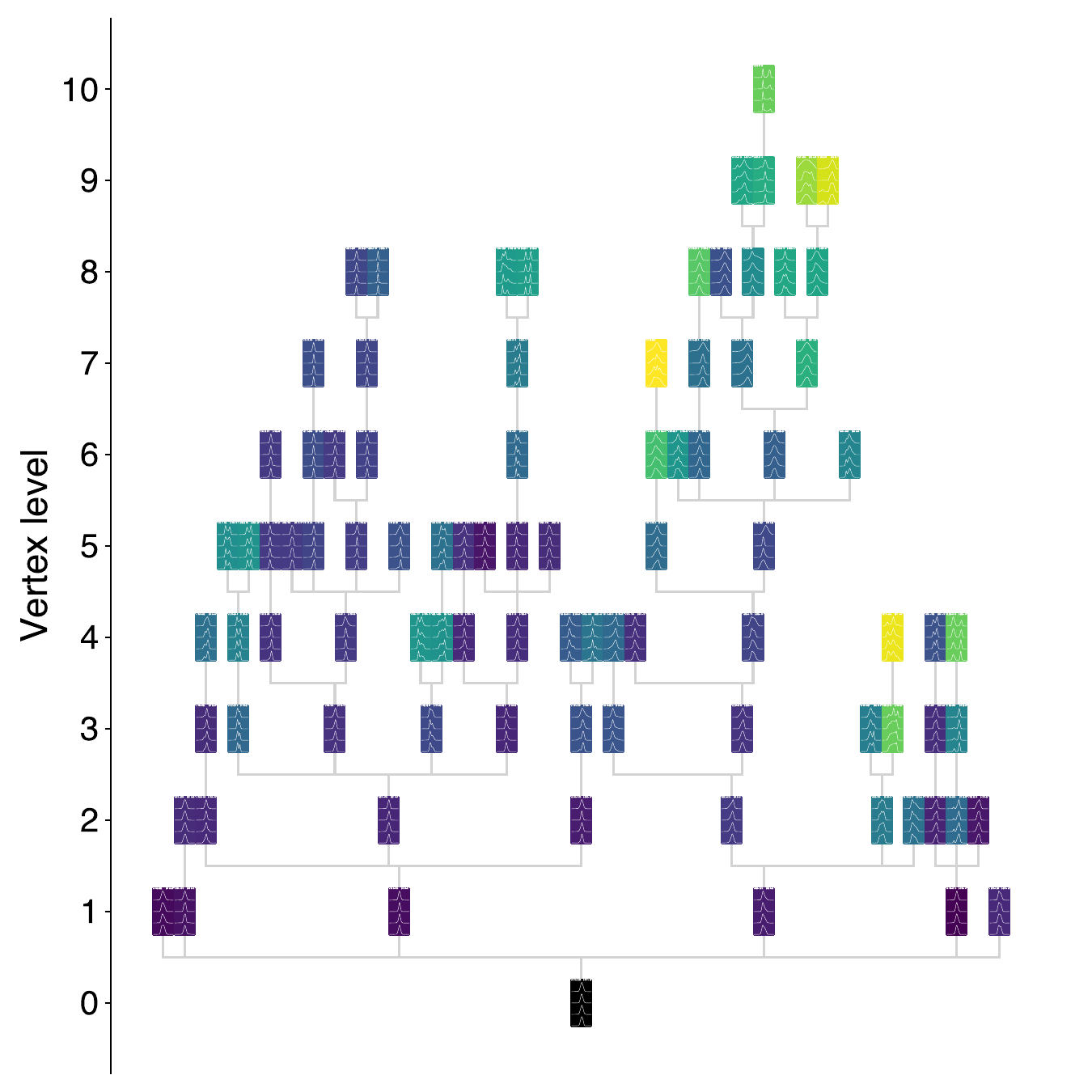} & 
\includegraphics[trim=0 0 0 0, clip, width=\columnwidth]{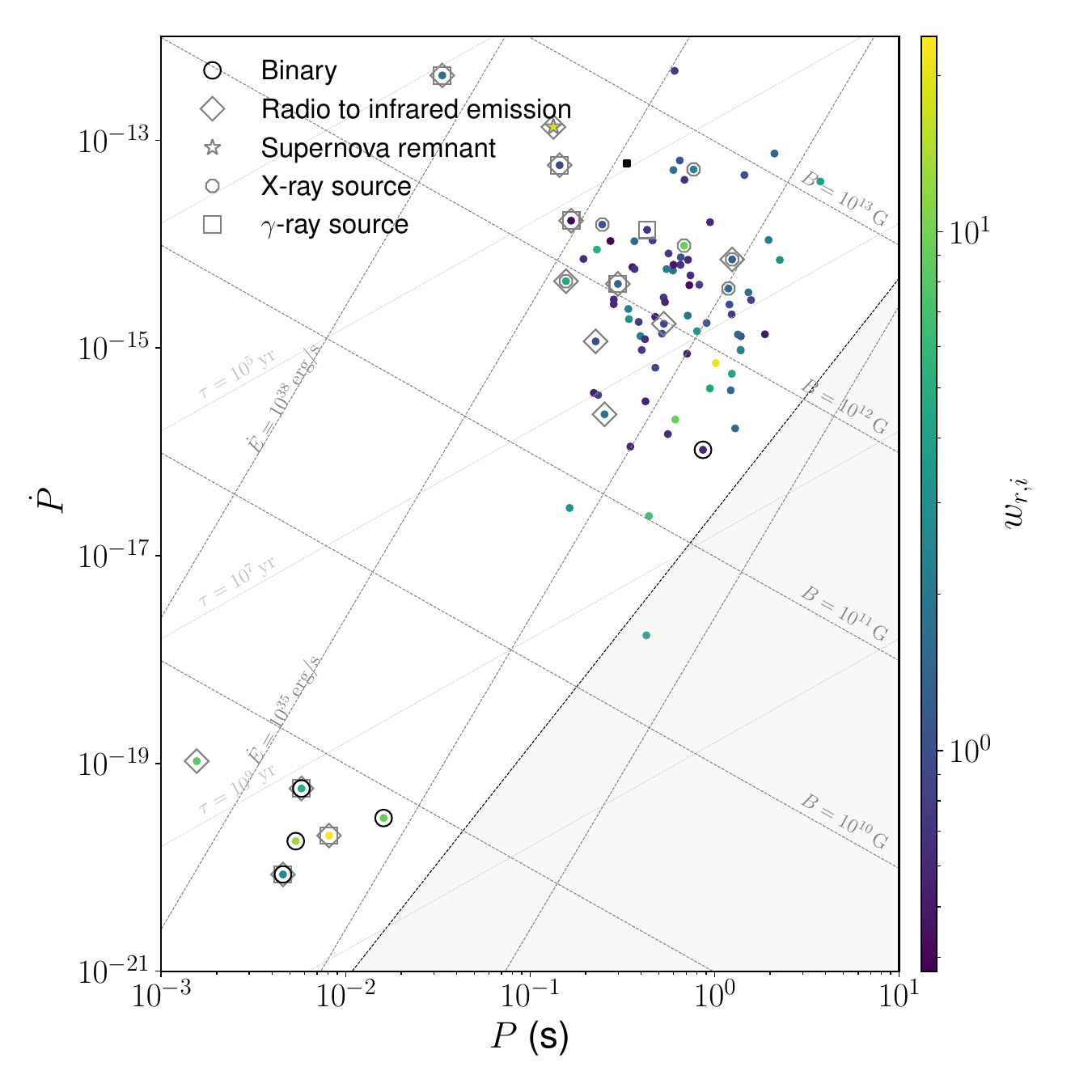}
\end{tblr}
\caption{MST (left) and $P-\dot{P}$ diagram (right) coloured with DTW distance $w_{r,i}$ between a given pulsar $i$ and the root pulsar $r$.}
\label{fig:p-pdot-w_r_i}
\end{figure*}

\paragraph{Branch 3} (42 vertices) starts with vertex pulsar J1903+0135, whose relatively simple profiles that 
narrow and widen as we move from high to low frequencies. 
At level 2, the rightmost short branch shows a similar wide to narrow to wide evolution, with the addition of outriders that are more prominent in the highest frequencies.
As we move up this branch, one of the two outrider components becomes more prominent than the other, and becomes more and more resolved at higher frequencies for the leaf (J1847$-$0402). 
The level 2 leftmost branch instead shows  is similar to this last branch, with the subtle difference of showing signs of more than one component within the outrider components (e.g. J2002+4050 and J0332+5434). While these are classified as T, one could equally argue that these are M.  

The main central branch at level 2 further splits into four sub-branches of diverse variation patterns at level 3. 
In this branch, the general trend seems to be a bright narrow component of constant (relative) amplitude, with one or more components gaining amplitude towards lower frequencies.
The level 3 leftmost branch morphs into M profiles, with two outer components, and a central component of lesser amplitude, with its amplitude increasing towards lower frequencies. 
The next branch from the left contains single profiles, including ${\rm S_t}$ and ${\rm S_d}$.
The third branch only contains T and M profiles, with one stable component and other varying over frequencies.

\paragraph{Branch 4} (33 vertices) begins with J0629+2415, a pulsar that shows an asymmetric profile at all frequency bins, with a component fading towards lower frequency. 
Pulsars following the main branch that is part of the longest sequence tend to conserve this general trend, although sometimes inverted in phase. 
Still in this branch, morphology begin to vary sharply at level 4 (J2326+6113). From there, sub-tracks form.
The leftmost branch of level 5 onwards is inhabited by wide, multi-component profiles, with components being more or less
resolved at varying frequency bins.
It is interesting to note that
while J0814+7429 (B0809+74)
is classified as a conal single
in the literature, 
more components are visible by eye. This agrees with the ${D \to S_d}$ track of B0809+74 in \citet{Rankin1983ApJ...274..333R},
and much of the discussion on this pulsar since \citep[e.g.,][]{lkr+02}. 
Here we thus have a number of profiles
where a conal single profile at one frequency evolves towards either a triple or multiple. 
The rightmost branch of level 5 has the pulsar J1932+2220 with one of the most peculiar morphology evolution in the set, with profiles widths varying from fairly wide at our highest frequency bin, then narrowing until the 700--1000\,MHz bin, and then widening again at the lowest frequency bin.
Potentially we here see a transition from intervening  broadening (dispersion, scattering) at lower frequencies --
J1932+2220 is relatively high DM -- to intrinsic broadening at higher frequencies.
This evolution pattern is only found again in J0534+2200. 
Other pulsars in this branch form a mix of fairly complex profiles evolution over frequency, and include all the
recycled pulsars in our sample (green boxes in the lower panel of Figure \ref{fig:tree-4freqs-w10-colour}).
Profiles show individual components amplitudes increasing towards higher frequencies. 
Moreover, a number of cases have components shifting in phase, alike aberration and retardation~\citep{GG2003ApJ...584..418G}.

Finally, profiles in the rightmost branch starting at level 2 have components being resolved at the lowest frequencies (top of a panel), which blend at higher frequencies, perhaps due to extra components becoming apparent and creating a plateau (as most obviously seen in J1900$-$2600). 

\paragraph{Branch 5} (eight vertices) starts with J0139+5814 which is formed of single profiles at all bins, with minor signs of an extra components at higher frequencies. It also broadens towards lower frequencies.
Pulsars in this branch also show an asymmetry in their main component, and generally bifurcate towards having at least two distinct components at lower frequencies, with the faintest component gaining in amplitude as we move towards higher frequencies. 
Here again, components tend to shift in phase as a function of frequency.

\begin{figure*}[th!]
\centering
\includegraphics[trim=0 0 0 0, clip, width=130mm]{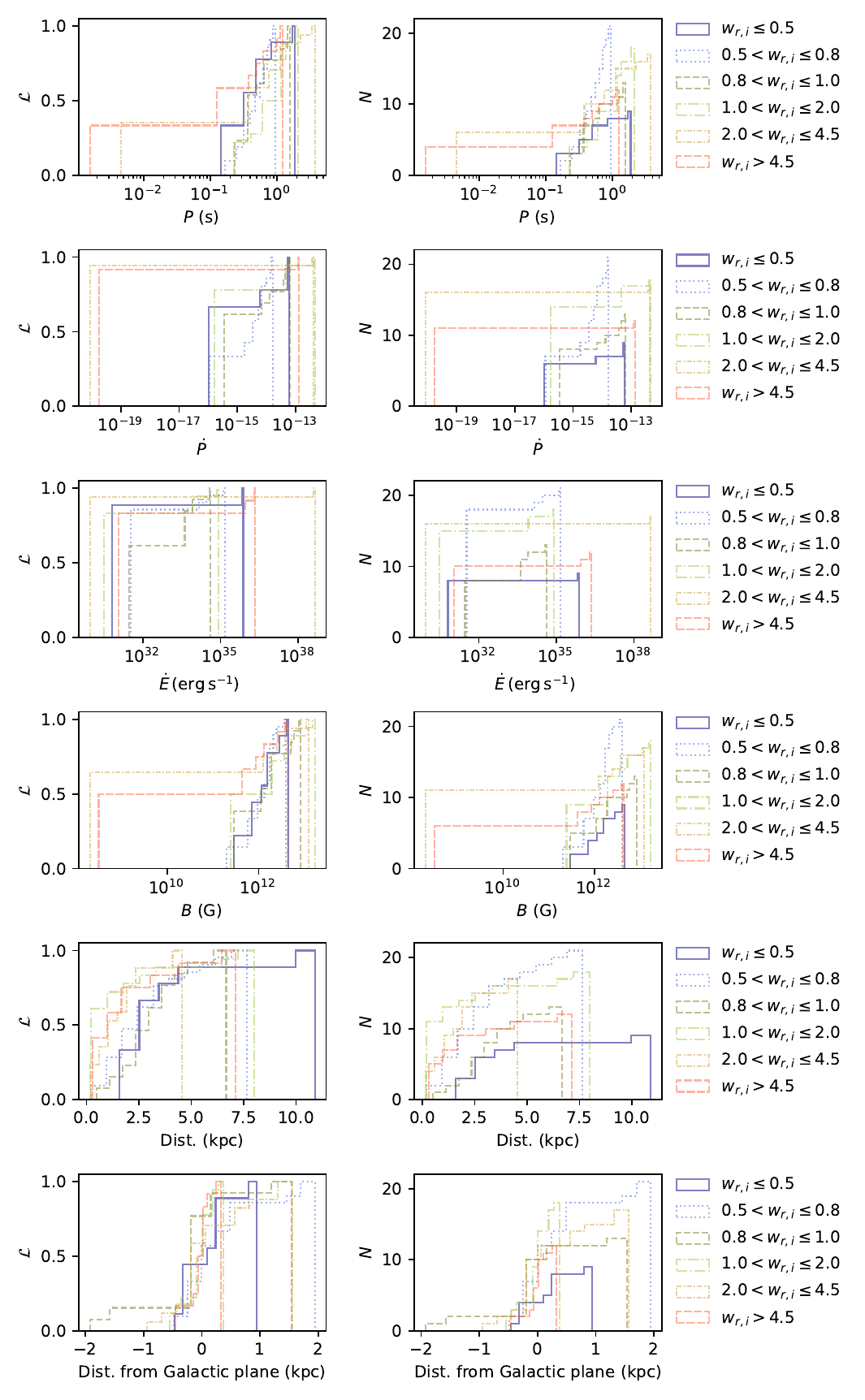}
\caption{Cumulative distributions (left: likelihood; right: counts) of physical parameters as a function of profile complexity ($w_{r,i}$,  Figure \ref{fig:p-pdot-w_r_i}). All values were taken from the ATNF pulsar catalogue using parameters P0, P1, EDOT, BSURF, Dist, and ZZ, respectively.}
\label{fig:cumulative-dists}%
\end{figure*}

Finally, we can utilise the MST to estimate morphological complexity since the root is occupied by a simple single profile, and leaves tend to be occupied by more complex profiles.
In Figure \ref{fig:p-pdot-w_r_i}, we colour each point by the DTW weight $w_{r,i}$ between the root $v_r$ and any other vertex $v_i$. 
Here, pulsars with $w_{r,i} \lesssim 0.5$ are generally composed of single-component profiles. 
At higher $w_{r,i} \gtrsim 0.5$, profiles become more complex.
The most simple profiles lie in the slow pulsar cloud ($P>0.1{\rm s}$, $10^{-16} \lesssim \dot{P} \lesssim 10^{-13}$). 
More complex profiles $w_{r,i} \gtrsim 1.5$ are all found at extremes (low and high $\dot{P}$, fast $P$). 
Pulsars with $w_{r_i} \gtrsim 2$ are found at all $P$, but are found below $\dot{E}\sim10^{35}\,{\rm erg\,s^{-1}}$.
We show cumulative distributions of physical parameters as a function of $w_{r, i}$ ranges in Figure \ref{fig:cumulative-dists}. 
There are hints that more complex profiles tend to lower $\dot{E}$ before diverging towards the millisecond period regime, which is in agreement with \citetalias{KJ2007}. 
Slightly more noticeable even in Figure~\ref{fig:p-pdot-w_r_i} is that profile complexity $w_{r,i}$
increases with characteristic age $\tau$. 
The Pearson correlation coefficient between $w_{r,i}$ and $\tau$ between is 0.34, not negligible, but low.

\section{Classifying new profiles from nearest neighbours}
\label{sec:classification}

\subsection{Motivation and methodology}
\label{subsec:classification-method}

Previously, newly discovered pulsars would generally be classified by eye
into the \citeauthor{Rankin1983ApJ...274..333R} scheme as one of core single (${\rm S_t}$), conal single (${\rm S_d}$), double (D), triple (T), or five-component/multiple (M).
In this scenario, an astronomer compares the profile and behaviour of the new source to that of known sources.
In this section, we investigate the case of automating  --  and potentially improving and unbiasing  --  the human-based classification through the use of the MST.
In particular, each source in our sample that is not yet categorised in the literature is compared to its nearest neighbours in the tree. 
From that we derive its class.

To carry out the experiment, we set the following rules. 
For each unclassified pulsar, we: 
(1) Set its class given the majority class of its direct neighbours; (2) if there is a tie among neighbours, assign the class of the neighbour with the smallest $w$; and
(3) if there remain unclassified pulsars after having visited all vertices in the MST, repeat steps $1$ and $2$, taking into account the newly predicted classes, until all pulsars have been classified.

As a test case, we apply this classification methodology to pulsars in our set that are unclassified in \citet{Rankin1990ApJ...352..247R},  \citet{2011ApJ...727...92M}.
\citet{Rankin1990ApJ...352..247R} classes are based on observations taken directly by the author at 325\,MHz, 1.4\,GHz, and 1.6\,GHz, and by \citet[][1.4\,GHz with Arecibo]{rsw89} plus \citet[][1.7\,GHz with Effelsberg]{xrss91}.
\citet{2011ApJ...727...92M} classes are based on observations taken at 325\,MHz and 1.4\,GHz.
As discussed earlier, we are aware that our current dataset necessarily consists of total-intensity profiles only,
while the quoted studies also incorporate polarisation information from bespoke, targeted observations.
We argue, however, that the current test case is the most appropriate first step for a population-based approach.
For instance, having profiles at four frequency bins should provide a proxy to components' width spectral evolution, which could in principle differentiate between ${\rm S_d}$ and ${\rm S_t}$.

Since a given profile morphology can evolve at different frequencies for a single pulsar, it is not straightforward to simply utilise automatically the classes gathered from the literature to our test set. 
As we mentioned in Section \ref{sec:intro}, Rankin hypothesised three typical evolution paths as a function both of radio frequency ($\nu_{\rm lower} \to \nu_{\rm higher}$) and orientation with respect to the line of sight: (1) ${(T?) \to \rm D \to S_d}$; (2) ${\rm S_t \to T \to D}$; and (3) ${\rm S_t \to T \to M \to (Q?)}$.
There is, however, no existing human-labelled classification of pulsars on such tracks; only in individual classes.
Therefore, the classes we use for this exercise are necessarily partial and not necessarily representative of which track a profile falls into. 
Moreover, one can imagine that track 1 could be combined to either 2 or 3, lending only the two following tracks: (2\arcmin) ${\rm S_t \to T \to D \to S_d}$ and (3\arcmin) ${\rm S_t \to T \to M \to (Q?)}$, with track 2\arcmin\ having a more obtuse $\beta$ than that of track 3\arcmin. 
Given these limitations of using a fixed class for multi-frequency profiles, in addition to the classification obtained using MST shown in Figure \ref{fig:tree-4freqs-w10-colour}, we also computed classes on a per-frequency-bin basis\footnote{Resulting per-frequency-bin MSTs can be found in Appendix \ref{app:other-msts}.}. 
Results are presented and summarised in Table \ref{tab:classes}. 

\begin{table*}
\center
   \caption{Classification based on MST.}
   \begin{threeparttable}
    \begin{tabularx}{18cm}{XXcccccccc} 
\hline \hline
Name & & Prev. & FB1 & FB2 & FB3 & FB4  &  Maj. & MP  & M-F \\
(J2000.0) &  (B1950.0) & & & & & & &FB1 \hspace{6mm}FB2 \hspace{6mm}FB3 \hspace{6mm}FB4& (\S\ref{sec:results}) \\
\hline 
J0014+4746   & B0011+47   & --  & ${\rm S_d}$ & T           & ${\rm S_d}$ & ${\rm S_d}$ & (${\rm S_d}$) & \begin{minipage}{.3\textwidth} \includegraphics[width=\linewidth, height=6mm]{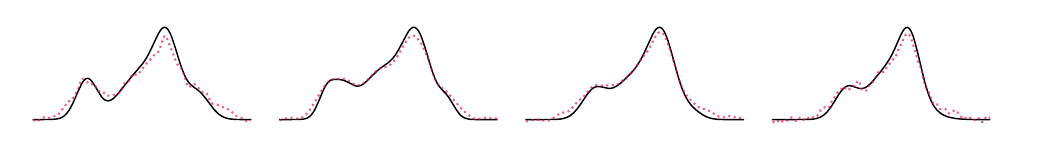} \end{minipage} & ${\rm S_d}$ \\
J0406+6138   & B0402+61   & --  & M           & M           & D           & T           & (M)           & \begin{minipage}{.3\textwidth} \includegraphics[width=\linewidth, height=6mm]{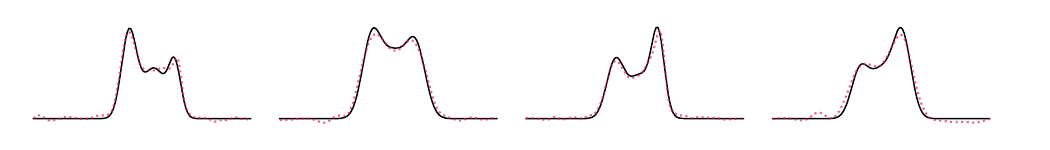} \end{minipage} & T           \\
J0437--4715 & --         & --  & ${\rm S_d}$ & ${\rm S_t}$ & ${\rm S_d}$ & ${\rm S_d}$ & (${\rm S_d}$) & \begin{minipage}{.3\textwidth} \includegraphics[width=\linewidth, height=6mm]{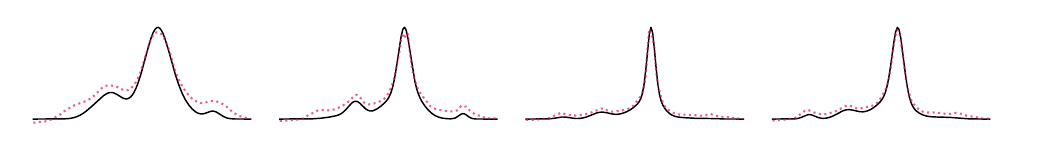} \end{minipage} & ${\rm S_d}$ \\
J0601--0527 & B0559--05 & --  & T           & ${\rm S_d}$ & M           & T           & (T)           & \begin{minipage}{.3\textwidth} \includegraphics[width=\linewidth, height=6mm]{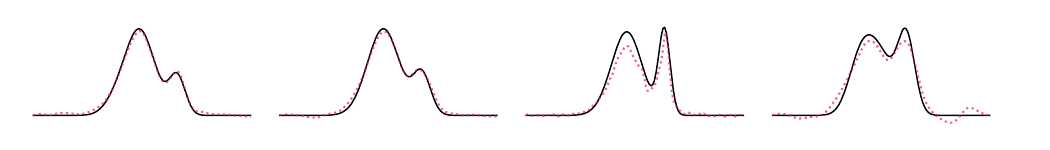} \end{minipage} & T           \\
J0953+0755   & B0950+08   & Sd? & ${\rm S_d}$ & ${\rm S_t}$ & ${\rm S_d}$ & ${\rm S_d}$ & (${\rm S_d}$) & \begin{minipage}{.3\textwidth} \includegraphics[width=\linewidth, height=6mm]{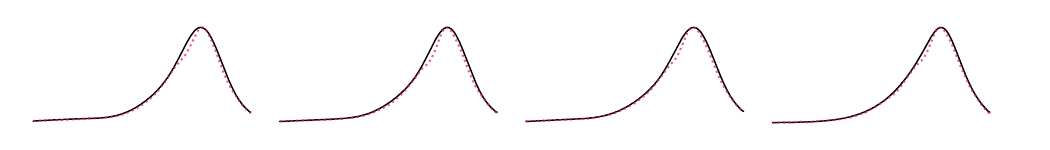} \end{minipage} & ${\rm S_d}$ \\
J1543--0620 & B1540--06 & Sd? & ${\rm S_d}$ & ${\rm S_t}$ & T           & ${\rm S_t}$ & (${\rm S_t}$) & \begin{minipage}{.3\textwidth} \includegraphics[width=\linewidth, height=6mm]{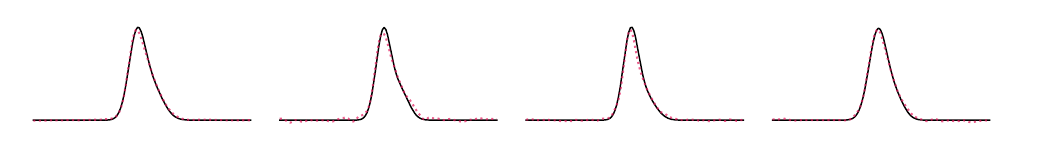} \end{minipage} & ${\rm S_t}$ \\
J1713+0747   & --         & --  & ${\rm S_d}$ & ${\rm S_t}$ & ${\rm S_d}$ & ${\rm S_d}$ & (${\rm S_d}$) & \begin{minipage}{.3\textwidth} \includegraphics[width=\linewidth, height=6mm]{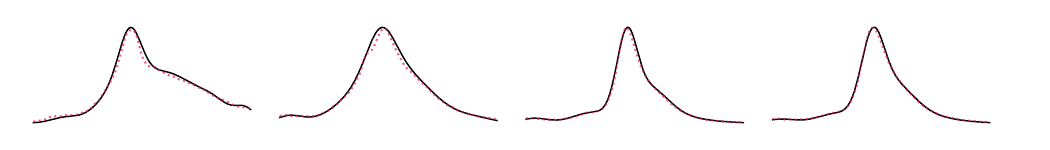} \end{minipage} & ${\rm S_d}$ \\
J1722--3207 & B1718--32 & --  & T           & M           & T           & ${\rm S_t}$ & (T)           & \begin{minipage}{.3\textwidth} \includegraphics[width=\linewidth, height=6mm]{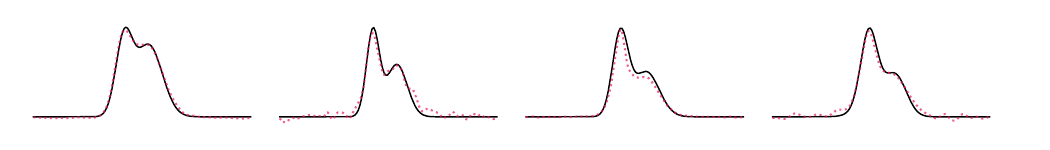} \end{minipage} & ${\rm S_t}$ \\
J1730--2304 & --         & --  & ${\rm S_d}$ & ${\rm S_t}$ & T           & ${\rm S_d}$ & (${\rm S_d}$) & \begin{minipage}{.3\textwidth} \includegraphics[width=\linewidth, height=6mm]{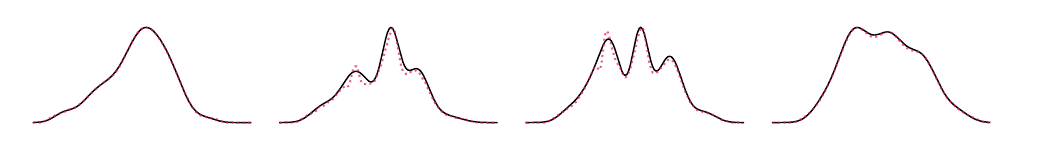} \end{minipage} & ${\rm S_d}$ \\
J1740--3015 & B1737--30 & --  & ${\rm S_t}$ & ${\rm S_t}$ & ${\rm S_d}$ & ${\rm S_t}$ & (${\rm S_t}$) & \begin{minipage}{.3\textwidth} \includegraphics[width=\linewidth, height=6mm]{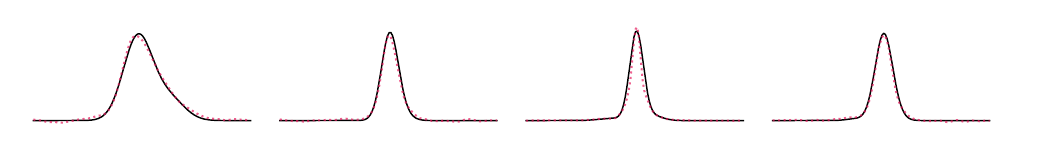} \end{minipage} & ${\rm S_t}$ \\
J1745--3040 & B1742--30 & T?  & ${\rm S_t}$ & T           & T           & T           & (T)           & \begin{minipage}{.3\textwidth} \includegraphics[width=\linewidth, height=6mm]{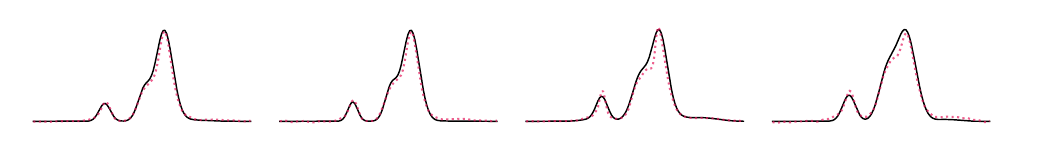} \end{minipage} & T           \\
J1803--2137 & B1800--21 & --  & ${\rm S_d}$ & ${\rm S_d}$ & ${\rm S_d}$ & ${\rm S_d}$ & (${\rm S_d}$) & \begin{minipage}{.3\textwidth} \includegraphics[width=\linewidth, height=6mm]{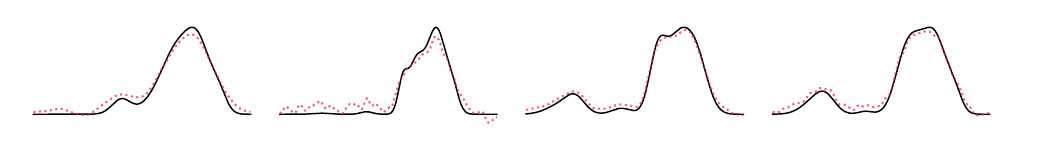} \end{minipage} & ${\rm S_d}$ \\
J1822--2256 & B1819--22 & --  & ${\rm S_t}$ & ${\rm S_d}$ & ${\rm S_t}$ & ${\rm S_t}$ & (${\rm S_t}$) & \begin{minipage}{.3\textwidth} \includegraphics[width=\linewidth, height=6mm]{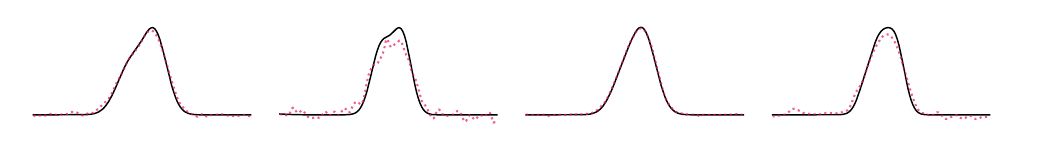} \end{minipage} & ${\rm S_t}$ \\
J1823--3106 & B1820--31 & --  & ${\rm S_t}$ & T           & T           & ${\rm S_d}$ & (T)           & \begin{minipage}{.3\textwidth} \includegraphics[width=\linewidth, height=6mm]{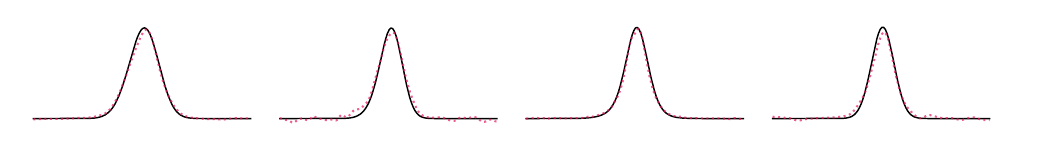} \end{minipage} & ${\rm S_d}$ \\
J1832--0827 & B1829--08 & --  & ${\rm S_t}$ & M           & M           & ${\rm S_t}$ & (--)          & \begin{minipage}{.3\textwidth} \includegraphics[width=\linewidth, height=6mm]{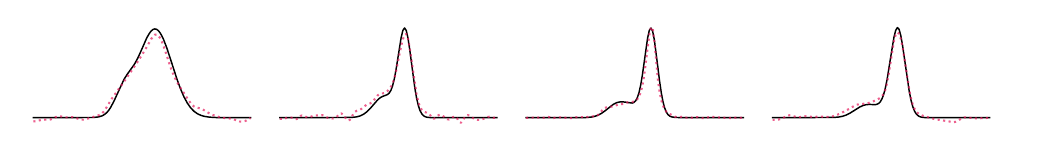} \end{minipage} & ${\rm S_t}$ \\
J1849--0636 & B1846--06 & --  & ${\rm S_t}$ & ${\rm S_d}$ & ${\rm S_d}$ & ${\rm S_d}$ & (${\rm S_d}$) & \begin{minipage}{.3\textwidth} \includegraphics[width=\linewidth, height=6mm]{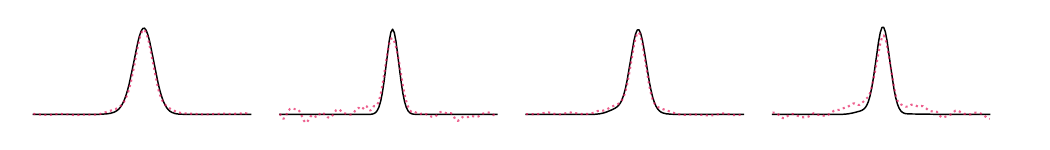} \end{minipage} & ${\rm S_d}$ \\
J1857+0943   & B1855+09   & T?  & ${\rm S_d}$ & ${\rm S_d}$ & ${\rm S_t}$ & ${\rm S_d}$ & (${\rm S_d}$) & \begin{minipage}{.3\textwidth} \includegraphics[width=\linewidth, height=6mm]{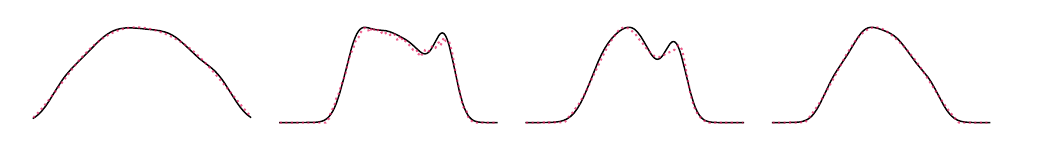} \end{minipage} & ${\rm S_d}$ \\
J1901+0331   & B1859+03   & --  & ${\rm S_d}$ & ${\rm S_d}$ & ${\rm S_t}$ & ${\rm S_t}$ & (--)          & \begin{minipage}{.3\textwidth} \includegraphics[width=\linewidth, height=6mm]{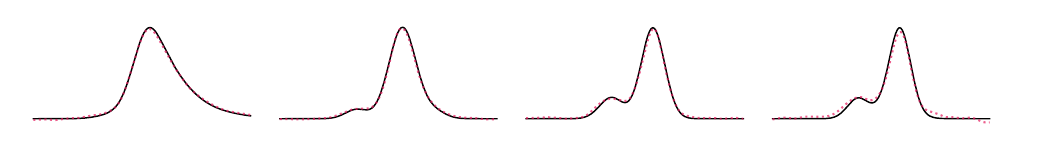} \end{minipage} & ${\rm S_t}$ \\
J1939+2134   & B1937+21   & --  & T           & T           & T           & T           & (T)           & \begin{minipage}{.3\textwidth} \includegraphics[width=\linewidth, height=6mm]{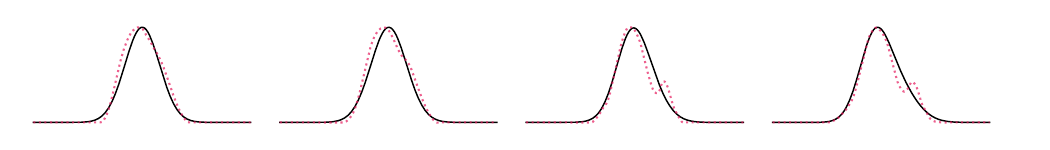} \end{minipage} & T           \\
J1946+1805   & B1944+17   & --  & T           & ${\rm S_d}$ & T           & ${\rm S_d}$ & (--)          & \begin{minipage}{.3\textwidth} \includegraphics[width=\linewidth, height=6mm]{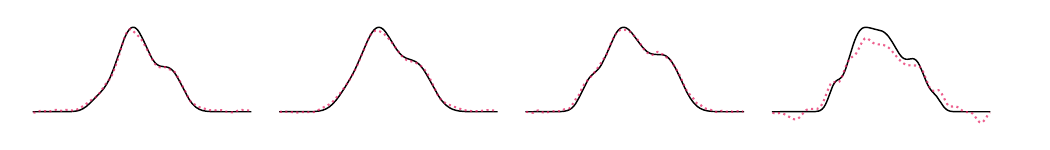} \end{minipage} & ${\rm S_d}$ \\
J1954+2923   & B1952+29   & --  & ${\rm S_t}$ & D           & T           & T           & (T)           & \begin{minipage}{.3\textwidth} \includegraphics[width=\linewidth, height=6mm]{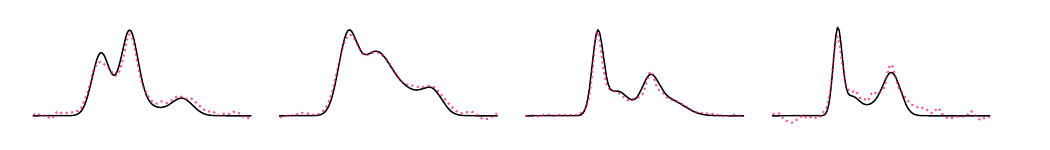} \end{minipage} & T           \\
J2002+4050   & B2000+40   & --  & D           & T           & ${\rm S_t}$ & ${\rm S_t}$ & (${\rm S_t}$) & \begin{minipage}{.3\textwidth} \includegraphics[width=\linewidth, height=6mm]{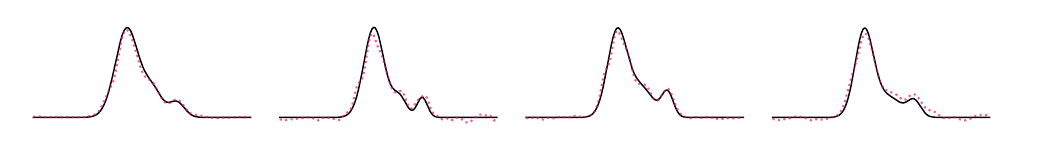} \end{minipage} & ${\rm S_t}$ \\
J2013+3845   & B2011+38   & --  & ${\rm S_d}$ & ${\rm S_d}$ & ${\rm S_d}$ & ${\rm S_d}$ & (${\rm S_d}$) & \begin{minipage}{.3\textwidth} \includegraphics[width=\linewidth, height=6mm]{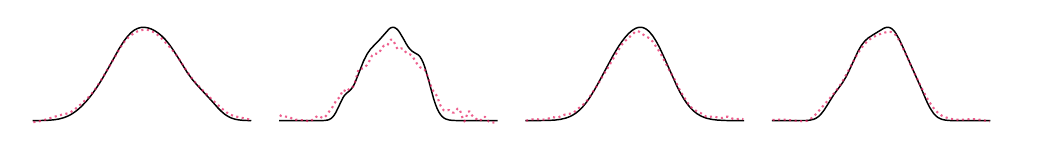} \end{minipage} & ${\rm S_d}$ \\
J2022+2854   & B2020+28   & T?  & T           & M           & M           & M           & (M)           & \begin{minipage}{.3\textwidth} \includegraphics[width=\linewidth, height=6mm]{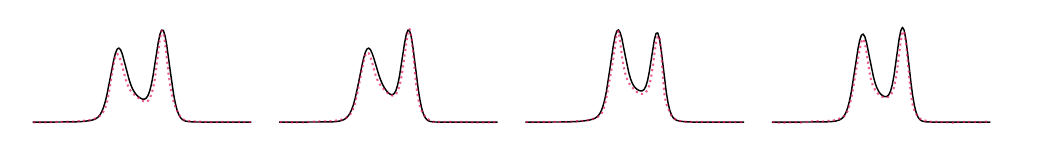} \end{minipage} & M           \\
J2145--0750 & --         & --  & D           & ${\rm S_t}$ & ${\rm S_d}$ & ${\rm S_d}$ & (${\rm S_d}$) & \begin{minipage}{.3\textwidth} \includegraphics[width=\linewidth, height=6mm]{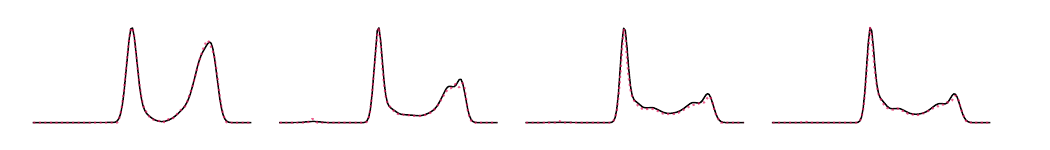} \end{minipage} & ${\rm S_d}$ \\
J2225+6535   & B2224+65   & --  & D           & D           & D           & T           & (D)           & \begin{minipage}{.3\textwidth} \includegraphics[width=\linewidth, height=6mm]{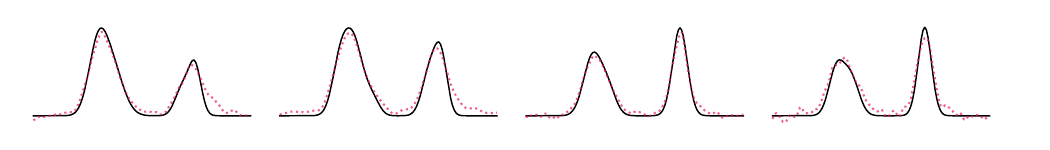} \end{minipage} & T           \\
J2313+4253   & B2310+42   & Sd? & ${\rm S_t}$ & M           & ${\rm S_t}$ & ${\rm S_t}$ & (${\rm S_t}$) & \begin{minipage}{.3\textwidth} \includegraphics[width=\linewidth, height=6mm]{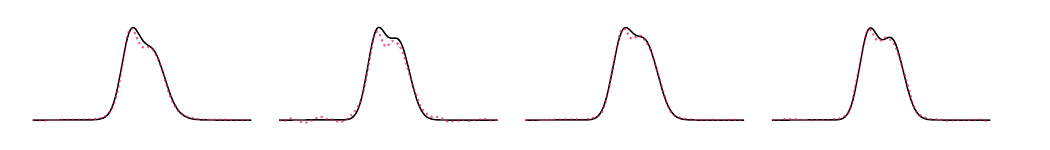} \end{minipage} & ${\rm S_t}$ \\
J2354+6155   & B2351+61   & --  & T           & T           & T           & D           & (T)           & \begin{minipage}{.3\textwidth} \includegraphics[width=\linewidth, height=6mm]{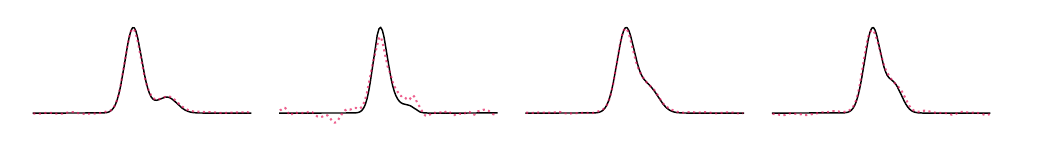} \end{minipage} & D           \\
\hline \hline 
    \end{tabularx}   
    \begin{tablenotes}
      \small
      \item Columns: (Prev.) Unclassified or ambiguous classification from our literature set. (FB1--4) Classes obtained using the MST and constructed using a single-frequency bin, namely (FB1) 400--700\,MHz, (FB2) 700--1000\,MHz, (FB3) 1000--1500\,MHz, (FB4) 1500--2000\,MHz. (Maj.) Majority based on FB1--4. (MP) Main pulse at each frequency bin, using the same phase width as Figure \ref{fig:tree-4freqs-w10}. (M-F) Class obtained using multi-frequency MST (Figure \ref{fig:tree-4freqs-w10-colour}).
    \end{tablenotes}    
    \end{threeparttable}
    \label{tab:classes}
\end{table*}

\subsection{Results}
\label{sec:classification-results}

We discuss our results in terms of internal consistency,
by contrasting the profiles' classification with our own qualitative classification opinions,
and by comparing against the literature. 

\subsubsection{General classification performance}

We first discuss the classes obtained on a single frequency bin, as shown in Table \ref{tab:classes}.
We see that in 25 out of 28 cases, a majority of  single-frequency classification agrees with one another.
We consider this 89\% agreement as solid evidence that the classification is consistent.
In half of the cases, the majority is 3 out of 4. 
In 3 cases it is unanimous. 
It appears the algorithm is robust in its classification. 

In most cases (20 out of 28), the classification obtained when using the average over all frequency bins
(\S\ref{sec:results}) in the construction of the MST agrees with the single-frequency MST majority.
We interpret this as further affirmation of the use of our algorithm.
We will discuss the outliers in more depth in the next subsection.

\subsubsection{Qualitative discussion of individual complex classification cases}

While the method generally agrees, three cases cannot reach a majority from their single-frequency classification,
as they are tied 2-2.
PSR~J1832--0827 is split between ${\rm S_t}$ and M, % not much lit on this psr B1829-08 (except the planet) -- JVL
J1901+0331 (B1859+03) is split between ${\rm S_t}$ and ${\rm S_d}$,
and J1946+1805 between ${\rm S_d}$ and T. 
With the exception of the classic indecisive single case that would require polarisation information, the other cases do show signs of both classes. 
For instance, J1832--0827 appears by eye as a partial cone (conal components only visible on one side of the core component),  with a sharp narrow component that is likely the core component, and leading conal outriders.  
We note that the difference between T and M isn't always obvious by eye, which may simply reflect the presence of blended components. 

In a few cases, the multi-frequency classification differs from the single-frequency majority:
\paragraph{J0406+6138} has an M majority, and a T for the multi-frequency MST. In this case, individual classes vary between M, D and T. Therefore, it is likely that we are observing the pulsar at relatively high $\beta$ angle. This pulsar has an exceptionally steep spectral index \citep{Krishnakumar_2019}, which is generally associated with core emission \citep[see, e.g.,][]{Rankin1983ApJ...274..333R}.
\paragraph{J1722--3207} has a T majority, and is classified as ${\rm S_t}$ by the multi-frequency MST.  This is likely due to the clear presence of a leading narrow component that hardly vary over frequency, while trailing and blended conal components slowly fade as we get to higher frequencies. \citet{2019MNRAS.483.2778I} find a steep polarisation-angle (PA) swing for this source. That is generally suggestive of core emission, agreeing with the classification.
\paragraph{J1823--3106} has a T majority, and is classified as ${\rm S_d}$ by the multi-frequency MST. In this case, the triple class is hard to understand given the models appear as single peak profiles. A closer look at FB2 does show a leading faint conal component that likely gave a slight asymmetry to the global morphology, or simply scatter broadening. One could therefore argue that this profile is on the $T\to {\rm S_d}$ track. However, the steep PA swing in \citet{2019MNRAS.483.2778I} does not immediately strengthen that case. 
\paragraph{J2225+6535} has a D majority, and is classified as T by the multi-frequency MST. The well-separated components do appear as double components. However, the asymmetry of the components enables one to argue that extra components are present. There are here some discrepancy between the model and the data, and it is possible that with a better model, the profile would have been classified as M. \citet{2016MNRAS.462.4416W} find a flat PA.  
\paragraph{J2354+6155 (B2351+61)} has a T majority,  and is classified as D by the multi-frequency MST. Here again, the profile visually appears as a partial conal profile, with a main core component and one (or more, blended) trailing conal outrider(s).  From polarisation information, \citet{2004MNRAS.352..689K} indeed conclude the emission is conal.

\subsubsection{Core versus conal single}

In a number of cases where multiple peaks are visible to the naked eye (Figure~\ref{fig:tree-4freqs-w10-colour}), the classification we obtain tends towards one of the two single classes: wide profiles are often classified as conal single, while those with a narrow component are often classified as core single. 

It is well known that witnessing a single peak provides no help in differentiating between ${\rm S_t}$ and ${\rm S_d}$, either requiring observations over a wide frequency range to determine if a single profile developed outriders at high frequency or bifurcated at low frequency; or polarimetry to identify the characteristic circular polarisation frequently exhibited by the core species or linear polarisation associated with the conal variety~\citep{Rankin1990ApJ...352..247R}.
Therefore, it is expected that the classification performed in this work would confuse ${\rm S_t}$ and ${\rm S_d}$
in certain cases. 

For example, J1822--2256 (B1819$-$22; \citealt{dls72}) is classified as core single in Table \ref{tab:classes}. 
In their subpulse-drift paper \citet{2018MNRAS.476.1345B} find the source is consistent with being ${\rm S_d}$.
Their conclusion is based mostly on polarisation data.  
However, the zero-crossing in Stokes V\footnote{\href{https://psrweb.jb.man.ac.uk/epndb/\#jk17/J1823-3106/J1823-3106.1400MHz.psrfits}{EPN profile at 1400\,MHz, last visited 2 November 2023.}} indicates that the line of sight crosses the pole and, hence, that we likely are looking at a core component. 
This is not the case in profiles at all frequencies, and hence, the single flavours may vary between cone and core components at different frequencies. 

It may well be that characteristics such as the pulse width or asymmetry automatically enable one to differentiate them.
We note that in the multi-frequency MST, some unclassified pulsars are surrounded by classified pulsars, and their predicted classes appear to be sensible. 
This is the case for J1832$-$0827 (B1829$-$08) for example -- predicted as ${\rm S_t}$, % no lit on B1829-08
and shares many similarities with J1759$-$2205 that is also classified as ${\rm S_t}$.
That is interesting, because from sub-pulse drift or polarisation studies,
little could previously be determined about the geometry of J1832$-$0827 \citep[e.g.,][respectively]{2006A&A...445..243W,2023MNRAS.520.4801J}.

Lastly, a select few of unclassified pulsars were found in one sequence within the MST, all of which derive their predicted
conal-single ${\rm S_d}$ class from the same root source: J0630-2834.
This is series J0953+0755 (B0950+08)
$\to$ J1713+0747 % The MSP
$\to$ J0014+4746 (B0011+47) and
J0437--4715  % Another MSP
$\to$ J2145--0750. % Yet another one 
We note that B0950+08 was already tentatively classed as conal
in \citet{Lyne_Manchester_1988MNRAS.234..477L}, which is encouraging,
but the source is complicated, as discussed in \citet{2003MNRAS.344L..69K} and  \citet{2022A&A...658A.143B}, for example.
Three of the four subsequent sources in the sequence are millisecond pulsars (MSPs; see next section).
In those, classification is generally elusive \citep[as discussed in e.g.][]{Rankin_2017}.
For J0014+4746, \citet{2022MNRAS.514.3202R} tentatively concluded that the emission is conal, in agreement with our automatic classification.

Overall, Table \ref{tab:classes} displays both the successes and current limitations of our method.
The algorithm is generally self-consistent, and a large majority of its findings are corroborated by studies on the
individual sources at hand, as described above. 
The limitations primarily follow from our fixed seed classes, which are used to classify other pulsars.
Arguably this is a limitation of the classification scheme in general.
Ideally, our conclusions in the `M-F' column should be an evolution track, not single classification.
However, we do not have at this stage the tools to automatically recognise such a track;
and there are basically no systematic human-labelled data on these tracks either.
We will discuss these results in more detail in Sect.~\ref{sec:discussion},
with an outlook to further improve 
our method, in future work.

\subsubsection{The MSPs}
It is striking that \name clusters all MSPs together.
This is best visible as the green-coloured set in the top-right part of
Figure~\ref{fig:tree-4freqs-w10-colour} (bottom panel).
This was achieved purely from the profiles, without spin information.  
While we cannot rule out a chance alignment, given the relatively low number of profiles,
we hypothesise that the actual underlying reason is twofold, and as follows.
First, the profiles are placed in a part of the graph that is generally inhabited by core and cone singles.
None of the MSP profile resemble conal doubles, for example, or five-component profiles.
The sampling time of our MSP profiles was high enough to rule out sampling smearing as hiding these features.
Thus, the MSP profiles seem to actually be self-similar.
Secondly, the MSPs are grouped at high vertex levels.
While they are smoothly connected to the normal population, the distances between these profiles is generally large.
Jointly those two causes group the MSPs together.

The fact that there is a smooth transition between the regular pulsars and the MSPs signifies that one physical
processes is fundamental to both. Much single-pulse behaviour that was previously though to only apply to regular
pulsars has since been observed in MSPs too: subpulse-drift \citep{2003A&A...407..273E} and mode changing
\citep{Mahajan_2018}, for example. We argue that classification and interpretation schemes, too,
should strive to include both recycled and non-recycled pulsars. 

\section{General discussion and future work}
\label{sec:discussion}

We evaluated \name to automatically seek for morphological types,  their evolution over frequency, and their global classes. 
For the first time, we present here a visual summary of 90 pulsars in a condensed format. 
All previous publications to date, to best of our knowledge, presented pulsars individually. 
Beyond the effort to utilise Rankin classes and predict classes from unclassified pulsars (\S\ref{sec:classification}), our condensed, self-organised method makes visually intelligible a number of trends in spectral evolution that only become obvious when considering a collection of pulsars. 
Utilising multiple frequency naturally leads to an MST that is slightly more complex to interpret than in the
single-frequency-bin case.
In the multi-frequency case, while we see individual branches display similar profile types, e.g. symmetrical triple, we also find profiles being arranged by frequency evolution patterns.
For example, a few pulsars show narrowing of profiles in FB2 and/or FB3 range before widening at FB1 and FB4 -- the most obvious cases being J1932+2220 and J0534+2200. 
In the case of J1932+2220, all profiles were taken from \citet{gl98}. The dispersion smearing ranges and the pulse-widths for this pulsar reported in their paper at different frequency bands are listed in Table~\ref{tab:smear}: \\

\begin{table}[]
    \centering
    \begin{tabularx}{\columnwidth}{XXX} 
    \hline
    Freq. & Disp. smearing & W50 \\
    (MHz) & (\%) & (\%) \\
    \hline
    410   &   2.6--4.7 & 6.1 \\
    610   &   0.7--1.4 & 3.7 \\
    925   &   0.4      & 1.7 \\
    1410  &   0.5--2.3 & 2.2 \\
    1600  &   1.5--1.6 & 4.4 \\
    \hline
    \end{tabularx} 
    \caption{Dispersion smearing and pulse width at various frequencies from \citet{gl98} for PSR~J1932+2220.}
    \label{tab:smear}
\end{table}

\vspace{1em}
With the exception of the profile observed at 925 MHz, the dispersive smearing could be contributing around 50\% or more to the pulse width (which could be up to 100\% at 1410 MHz). 
Therefore, the observed intriguing trend of the pulse width over frequency for this pulsar might simply be determined by the instrumental setup, and not be intrinsic to the pulsar emission regions. Similar effects might also be present in other relatively high dispersion measure pulsars with intrinsically narrow profile widths.

The trends we see in the MST can be due to multiple factors.
Firstly are geometric parameters such as $\beta$ or $\zeta$. 
Clearly these can determine how many, if any, core and cone components cross our line of sight.
Secondly, there may be intrinsic sub-families of pulsars, based, for example, on relatively unchanging characteristics such
as the magnetic field strength $B$ (magnetars -- regular pulsars -- MSPs).
Finally, the state of the emitting system may evolve when the pulsar period decreases,
and this cosmic accelerator \citep{1975ApJ...196...51R,2012ApJ...752..155V} operates on an
ever lower spin-down energy budget.
In this case, the pair-formation cascade over the polar cap may be less efficient
at lower $\dot{E}$ \citep{2019ApJ...871...12T},
possibly somehow causing the change from core to conal dominated profiles \citep[see][for a
  discussion]{2022MNRAS.517.1189O}.

\begin{figure*}[h!]
\centering
\begin{tabular}{cccc}
\multicolumn{2}{c|}{Case 1}&\multicolumn{2}{|c}{Case 2} 
\\ 
\hline \\
\includegraphics[trim=0 0 0 0, clip, width=3.4cm]{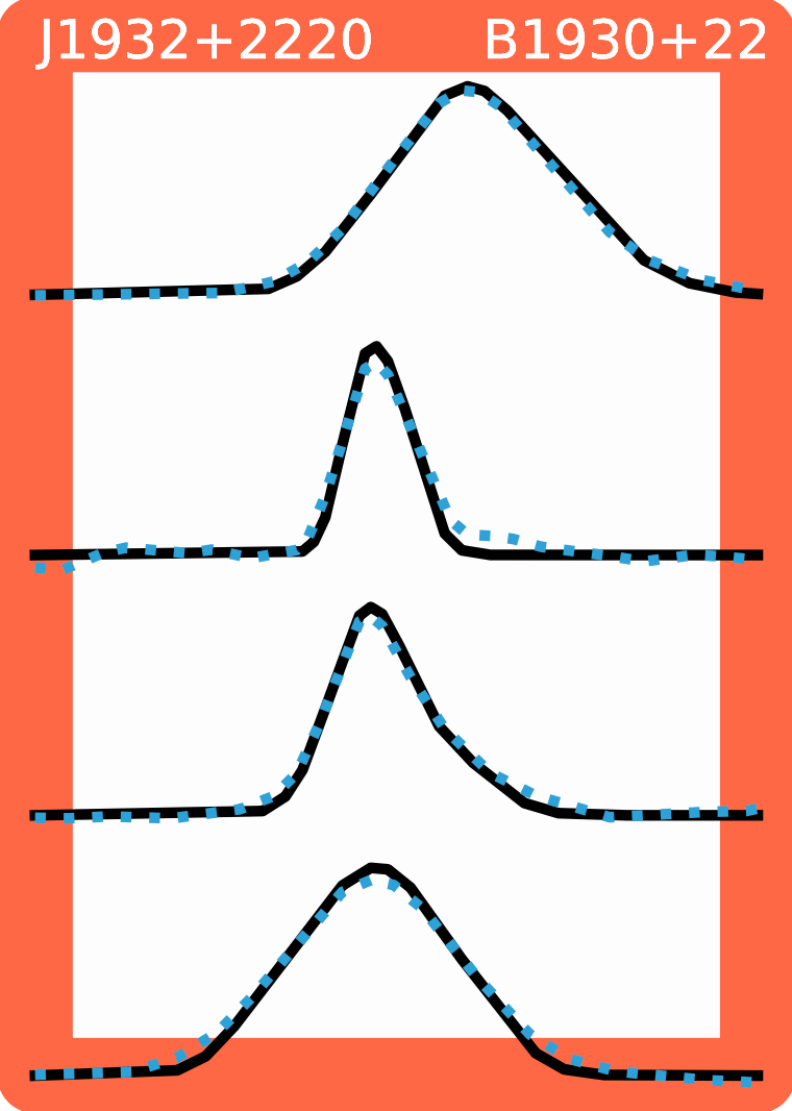} & 
\includegraphics[trim=0 0 0 0, clip, width=3.4cm]{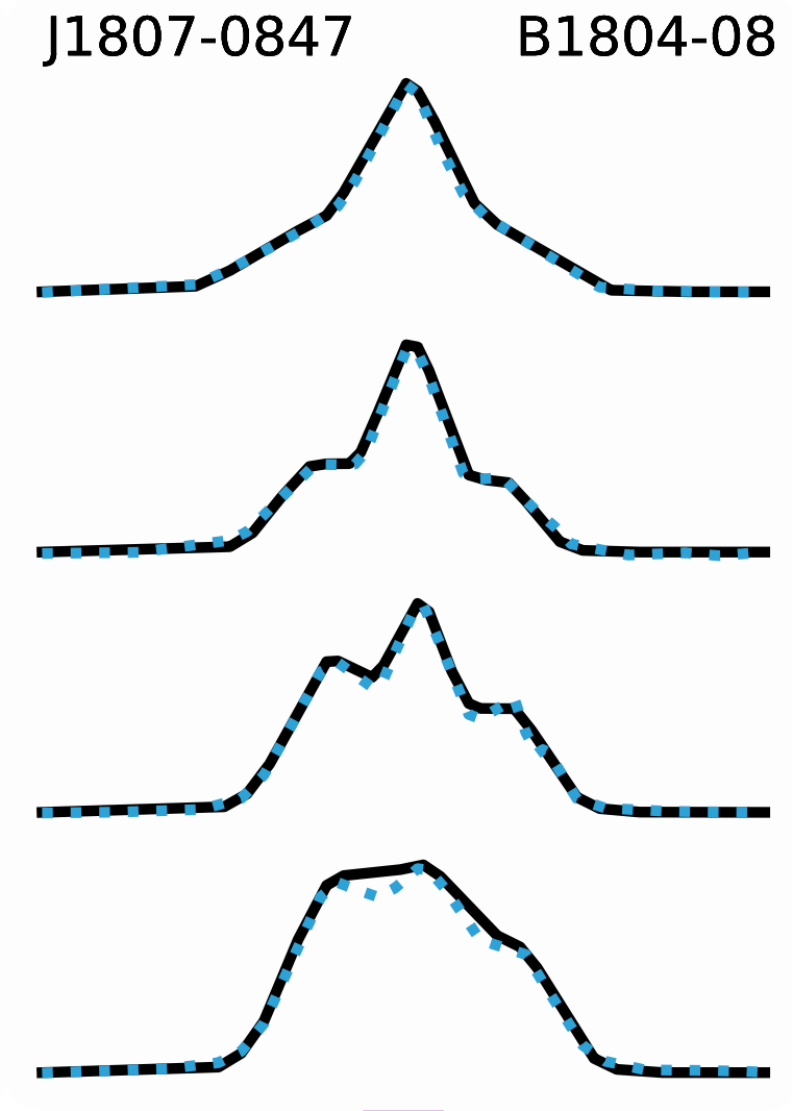} &
\includegraphics[trim=0 0 0 0, clip, width=3.4cm]{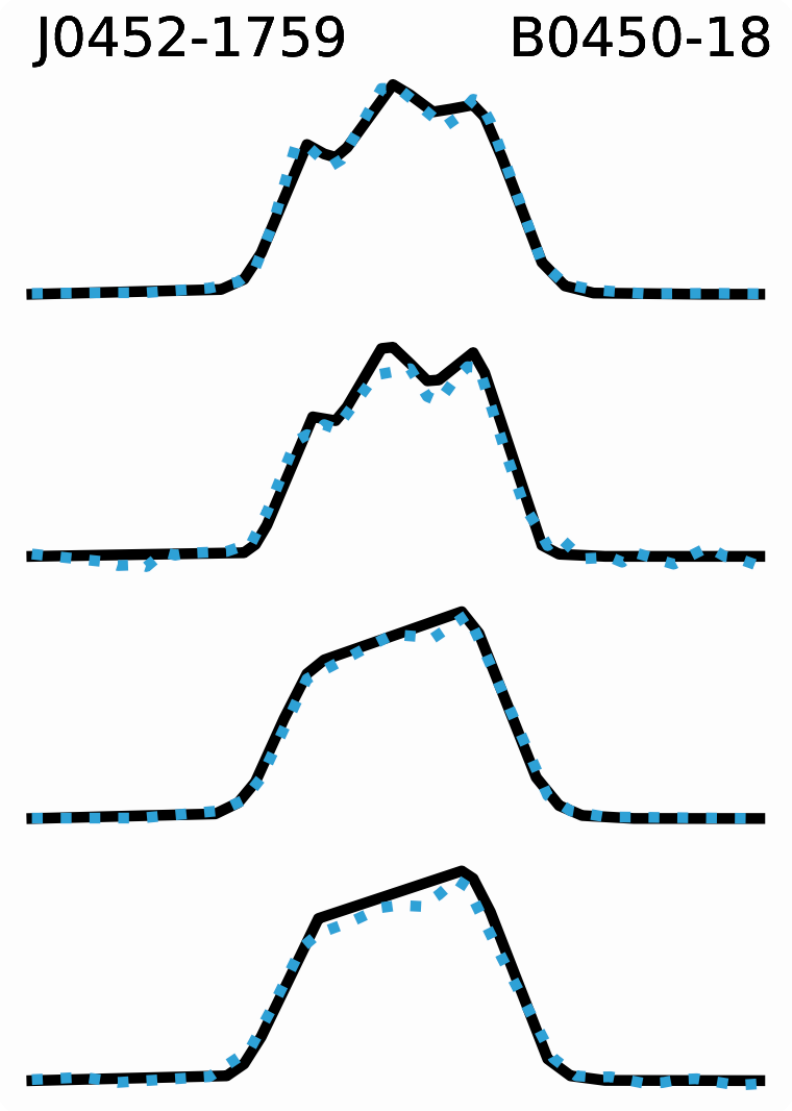} &
\includegraphics[trim=0 0 0 0, clip, width=3.4cm]{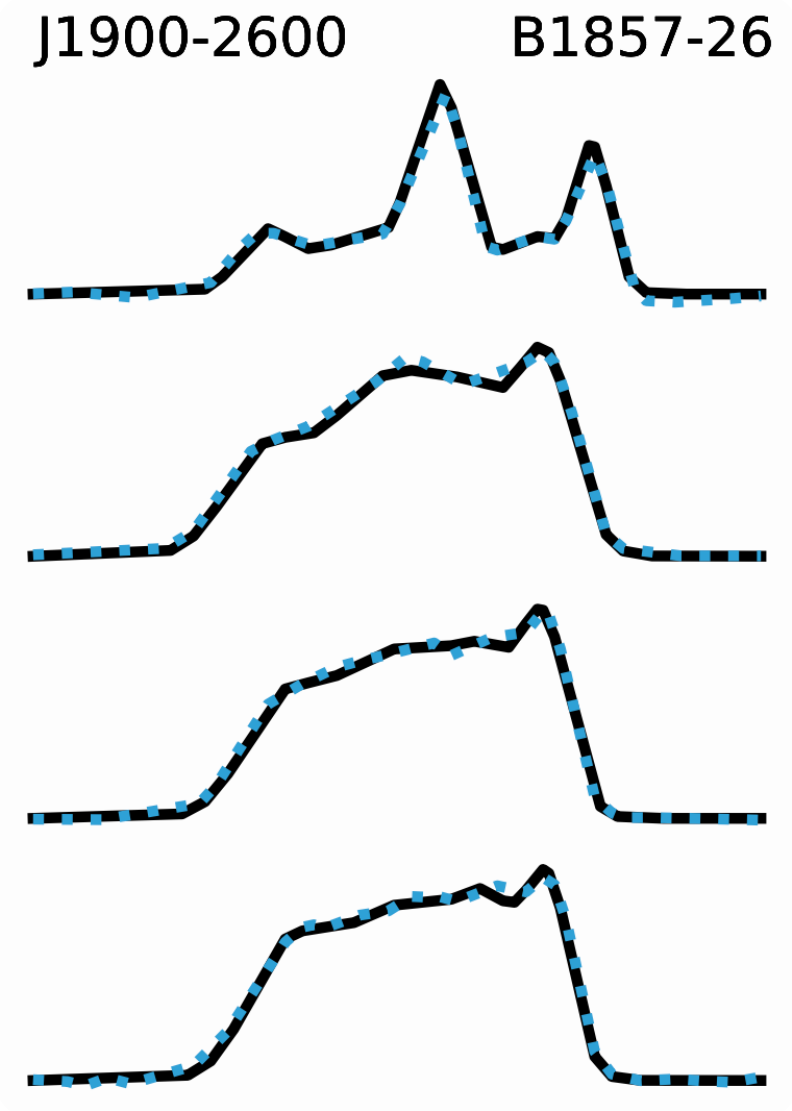}
\end{tabular}
\caption{Two example cases of neighbouring pulsars raising the question whether conal components are visible due to a geometrical effect or to the pulsar's state. Spin parameters were taken from the ATNF pulsar catalogue: \\
J1932+2220 (branch 4, level 5): $P$$\sim$$0.14$\,s, $\dot{P}$$\sim$$10^{-13.2}$, $\dot{E}$$\sim$$10^{35.9}$\,erg/s, $B$$\sim$$10^{12.5}$\,G; \\
J1807--0847  (branch 4, level 6): $P$$\sim$$0.16$\,s, $\dot{P}$$\sim$$10^{-16.5}$, $\dot{E}$$\sim$$10^{32.4}$\,erg/s, $B$$\sim$$10^{10.8}$\,G;\\
J0452--1759  (branch 4, level 3): $P$$\sim$$0.54$\,s, $\dot{P}$$\sim$$10^{-14.2}$, $\dot{E}$$\sim$$10^{33.1}$\,erg/s, $B$$\sim$$10^{12.3}$\,G;\\
J1900--2600 (branch 4, level 4): $P$$\sim$$0.61$\,s, $\dot{P}$$\sim$$10^{-15.7}$, $\dot{E}$$\sim$$10^{31.6}$\,erg/s, $B$$\sim$$10^{11.6}$\,G.
}
\label{fig:unresolved}%
\end{figure*}

These three general factors are not easy to disentangle. 
For example, if we compare the overall profile shape of J1932+2220 to that of its direct neighbour J1807--0847 in Figure \ref{fig:tree-4freqs-w10}, both having similar profile envelopes but slightly different periods, period derivatives, spin-down energies and magnetic field strengths. 
It is plausible that conal components similar to those observed in J1807--0847 are also present in J1932+2220 but blended in phase with the central component due to higher $B$, $\dot{E}$ and faster $P$ (Figure \ref{fig:unresolved}). 
A similar argument can be put forward for PSRs J0452$-$1759 and J1900$-$2600, which are also direct neighbours in the MST.
Future similarity investigations of profile shape and physical parameters that include polarisation information and high phase resolution should allow for further investigation into these questions.

% %% Pol, s.i.
In the work we present here, our distance measure does not depend on 
spectral index or the polarisation profile, fraction, and polarisation position angle. 
The strength of our approach is in the large number of profiles it compares. 
While the data provided by the EPN database forms the largest curated and publicly available set, the metadata of its profiles do not unambiguously state if the profile is flux and/or polarisation calibrated.
The data reduction underlying the submissions is quite heterogeneous, and profiles were submitted in different formats with different header requirements. 
We thus concluded that only flux-normalised Stokes-I profiles are robust enough at the moment for determining the distance between these profiles. 
Future application of our method on sets of more homogeneously analysed pulse profile collections such as the MeerKAT Thousand Pulsar Array~\citep{2020MNRAS.493.3608J} or Pulsar Radio Emission Statistics Survey (PRESS, PI: W. van Straten) with the ultra-wide bandwidth receiver at Parkes could benefit from also including spectral index or polarisation information.
We could then also extend our method to utilise frequency-dependent profile templates~\citep{Pennucci2019ApJ...871...34P} to compare pulsars rather than a select few frequency bins. 

ToPP is a method based in graph theory and optimisation methods, which are inherent concepts in artificial intelligence and machine learning. 
Given our specific interest in investigating the morphology of pulsar average profiles, we did not aim at curating the most generic method this present work. 
Rather, we tailored a method that allowed us to peer at features of interest with respect to common emission models such as the core-cone model in order to automatically organise them by morphology and spectral evolution.
We curated the algorithm with domain-knowledge (choice of shape-based metric, asymmetric step pattern, and symmetry constraint) rather than making it purely agnostic. 

The {\tt sequencer} algorithm presented by \citet{Baron+2021ApJ...916...91B} is fully agnostic, aggregating distances measured at different scales, and evaluating multiple distance metrics (L2, Kullback-Leibler Divergence, Monge-Wasserstein or Earth mover distance) to seek for the longest manifold. 
The {\tt sequencer} utilises the graph width and length to evaluate how much information is carried by the longest sequence. 
Branches outside the longest sequence are completely omitted as the algorithm specifically seek for scaling relations. 
ToPP extends this shortcoming through visualisation of the full MST, and the ability to investigate information from specific branches and sub-branches programmatically. 
For very large datasets, visualising the full tree may become impractical. 
To that end, edge contraction (defined in \S\ref{sec:graph}) and presented in \citet{vohl_2021_7704065b} can provide ways to synthesise the MST to its most informative vertices. 

% Other machine learning approaches could be considered to investigate pulsar profiles, such as self-organizing maps, 
\citet{Shan2015A&C....11...55S} considered the classification of single and double peak pulsar profiles by applying the fuzzy C-means clustering algorithm in conjunction with wavelet-based feature extraction, showing singularity-based features provided smaller classification error than shape-based extracted features. 
We note that our method does not specifically use extracted feature to classify profiles. 
Furthermore, fuzzy C-means clustering requires the number $C$ of clusters to be specified, which is limiting in cases where $C$ is not known a priori. 
However, this issue could be alleviated with a Dirichlet process approach. 
Future efforts to generalise ToPP should consider an evaluation of wavelet-based feature extraction as potential alternative to DPGMM prior to building the complete undirected weighted graph.

As opposed to other hierarchical clustering methods such decision trees, which require a training set, ToPP (DTW, DPGMM, MST) is inherently unsupervised. 
Many other unsupervised clustering methods exist -- each with pros and cons with respect to various aspects, such as time complexity, memory complexity, and the ability to deal with noise -- and we refer the interested reader to recent reviews~\citep{LABBE2023555, 2024arXiv240107389Y}.

The various MSTs presented above highlight ToPP's capacity to organise pulsars as a function of profile morphology (given DTW), leading to the classification results presented in \S\ref{sec:classification-results}. 
Beyond classification, one could utilise an MST to recover profiles with similar (global\footnote{Further extension of the algorithm could include means to fetch similar local features as well.}) morphological features as a pulsar of interest. 
This could be done by visiting close neighbours of the pulsar of interest, possibly limiting the search by setting a threshold on $w$. 
Future work could test such a scenario through a dedicated simulation study.

\section{Conclusion}
\label{sec:conclusion}

The integrated pulse profile of radio pulsars represents one of their most informative traits, carrying information on properties determined by time-stable factors such as the geometry or dominance of the strong magnetic field.
Pulsar profiles have enabled theorists to model the configuration of its magnetosphere and emission mechanisms. 
However, linking the pulsar radio emission properties to its identity and geometry remains difficult. 
We have presented {\name}, a methodology to automatically compare a population of pulsar average profiles and organise them by similarity. 
By comparing the shape of modelled profiles, we constructed MSTs, yielding branches inhabited by pulsars displaying similar morphologies and frequency evolution and evolving from simple cases near the MST root towards more complex features near the leaves. 

Employing the MST, we investigated the case of predicting the `Rankin class' based on a pulsar's location within the MST. 
We showed that the MST can distinguish, in many cases, between core and conal single profiles without any knowledge of spectral index or any polarisation information. 
We found limitations in using a fixed class when considering profiles over a range of frequencies. 
Future work should consider using morphological tracks such as the tree tracks introduced by \citet{Rankin1983ApJ...274..333R} as labels when classifying pulsars on multi-frequency MST. 
Furthermore, future application of our method on sets of more homogeneously gathered and analysed pulse profile collections would allow the polarisation information to be used within the similarity measure.

Given that the MST root vertex (in a sense, the average profile) is composed of a single profile at all of our frequency bins, with little to no frequency evolution, we utilised the root to evaluate profile complexity.
After evaluating the link between profile morphological complexity and pulsar identity ($P$, $\dot{P}$, $B$, $\dot{E}$; Figure \ref{fig:p-pdot-w_r_i} and \ref{fig:cumulative-dists}), we noted that the more complex profiles tend to a lower $\dot{E}$, in line with results from \citetalias{KJ2007},
and a higher characteristic age $\tau$.
Overall, the succinct MST provides a way to identify frequency evolution trends, and 
these trends can be of geometric or energetic nature.

With \name, we have only just started to scratch the surface of the condensed information provided by graph
representations.
As the number of known pulsars grows into the many thousands, such structured representations are essential for both
finding the population trends and identifying interesting outliers.

\begin{acknowledgements}
DV acknowledges funding from the Netherlands eScience Center grant AA-ALERT (027.015.G09).
JvL and YM were supported by 
the European Research Council (ERC) under the European Union's Seventh Framework Programme
(FP/2007-2013)/ERC Grant Agreement No. 617199 (`ALERT').
JvL further acknowledges funding from 
Vici research project `ARGO' (number 639.043.815),
and from CORTEX (NWA.1160.18.316), under the research programme NWA-ORC,
both financed by the Dutch Research Council (NWO).
We thank Michael Keith for help with accessing the EPN database.
DV thanks Anna Bilous, Jason Hessels, Simon Johnston, Aditya Parthasarathy, Andrzej Szary, and Joel Weisberg for useful discussions.
In memory of Fran\c cois Laviolette. 
Software: Astropy~\citep{astropy:2013, astropy:2018, astropy:2022} version 5.2, dtw-python~\cite{giorgino2009computing} version 1.3.1, Matplotlib~\citep{Hunter:2007} version 3.6.2, Numpy~\citep{harris2020array} version 1.24.1, Scipy~\citep{2020SciPy-NMeth} version 1.10.0, Pandas~\citep{reback2020pandas} version 1.5.2, psrqpy~\citep{psrqpy} version 1.2.4, scikit-learn~\citep{scikit-learn} version 1.2.0. 
\end{acknowledgements}
%%%%%%%%%%%%%%%%%%%%%%%%%%%%%%%%%%%%%%%%%%%%%%%%%%

%%%%%%%%%%%%%%%%%%%% REFERENCES %%%%%%%%%%%%%%%%%%

%\bibliographystyle{aa}
\bibliographystyle{yahapj}  % has clickable links
\bibliography{bibliography,journals,psrrefs}{}

\onecolumn 
\begin{appendix}
\section{Pulsar set}
\label{app::pulsar_set}

\begin{figure}[H]
\centering
\includegraphics[width=8.5cm]{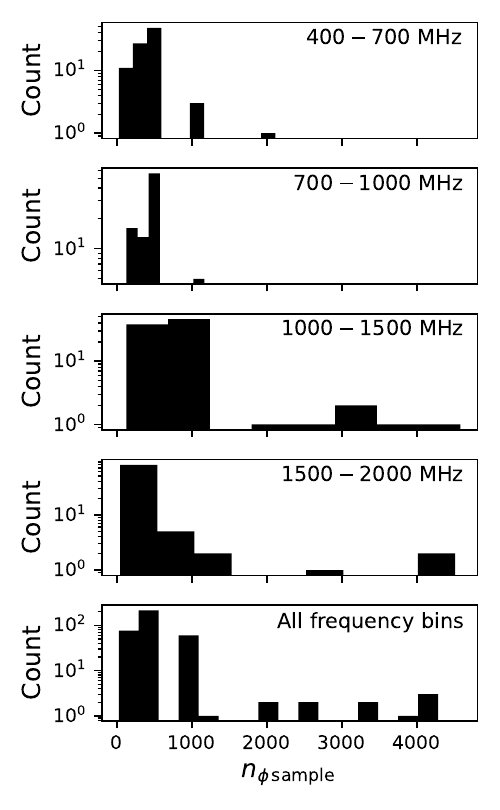}
\caption{Distribution of phase sampling rate (n) among profiles in our subset population prior to re-gridding to a common $n=512$ bins through spline interpolation. Each histogram constructed using Scott's rule. See text for details on sample selection and preparation.}%
\label{fig:epn-bins}%
\end{figure}

This section presents information about the pulsar set utilised in our study, including pulsar names, physical parameters, and the original reference for each profile (Table \ref{tab:pulsars}).
Physical parameters and uncertainties were taken from the ATNF pulsar catalogue~\citep{Manchester+2005AJ....129.1993M}.
Distances in the ATNF pulsar catalogue are based on the YMW16 electron-density model~\citep{Yao2017ApJ...835...29Y}; with relative errors within the uncertainty limits of independently measured distances on which the model is based. While relative errors have a factor of 0.9 as an upper limit, the mean and median have factors of 0.42 and 0.05 respectively. Hence, we choose to use a fiducial factor of 0.25.
In addition, we present the original phase sampling distributions in Figure \ref{fig:epn-bins}.

\begin{longtable}{llrrrrrrrr} 
   \caption{Pulsars, physical parameters, and EPN references.\label{tab:pulsars}}\\

\hline\hline
Name & & $P$ & $\log\dot{P}$ & $\log B$ & $\log\dot{E}$ & $\log{\tau_c}$ & Dist. & Dist.${\rm _{GP}}$& References \\
(J2000.0) &  (B1950.0) & (s) & & (G) & (erg\,s$^{-1}$) & (yr) & (kpc) & (kpc) &  \\
\hline 

$J0014+4746$ & $B0011+47$ & $1.241$ & $-15.2$ & $11.9$ & $31.1$ & $7.5$ & $1.776\pm0.444$ & $-0.419\pm0.105$ & a, d, d, a \\
$J0034-0721$ & $B0031-07$ & $0.943$ & $-15.4$ & $11.8$ & $31.3$ & $7.6$ & $1.030\pm0.258$ & $-0.939\pm0.235$ & a, a, b, g \\
$J0139+5814$ & $B0136+57$ & $0.272$ & $-14.0$ & $12.2$ & $34.3$ & $5.6$ & $2.600\pm0.650$ & $-0.151\pm0.038$ & a, a, a, a \\
$J0141+6009$ & $B0138+59$ & $1.223$ & $-15.4$ & $11.8$ & $30.9$ & $7.7$ & $2.300\pm0.575$ & $-0.053\pm0.013$ & a, a, a, a \\
$J0304+1932$ & $B0301+19$ & $1.388$ & $-14.9$ & $12.1$ & $31.3$ & $7.2$ & $0.739\pm0.185$ & $-0.377\pm0.094$ & a, a, c, d \\
$J0332+5434$ & $B0329+54$ & $0.715$ & $-14.7$ & $12.1$ & $32.3$ & $6.7$ & $1.695\pm0.424$ & $-0.005\pm0.001$ & a, a, e, a \\
$J0358+5413$ & $B0355+54$ & $0.156$ & $-14.4$ & $11.9$ & $34.7$ & $5.8$ & $1.000\pm0.250$ & $0.044\pm0.011$ & a, d, a, a \\
$J0406+6138$ & $B0402+61$ & $0.595$ & $-14.3$ & $12.3$ & $33.0$ & $6.2$ & $4.545\pm1.136$ & $0.596\pm0.149$ & a, d, a, d \\
$J0437-4715$ & -- & $0.006$ & $-19.2$ & $8.8$ & $34.1$ & $9.2$ & $0.157\pm0.039$ & $-0.078\pm0.019$ & k, f, f, k \\
$J0452-1759$ & $B0450-18$ & $0.549$ & $-14.2$ & $12.3$ & $33.1$ & $6.2$ & $0.400\pm0.100$ & $-0.196\pm0.049$ & a, a, b, a \\
$J0454+5543$ & $B0450+55$ & $0.341$ & $-14.6$ & $12.0$ & $33.4$ & $6.4$ & $1.180\pm0.295$ & $0.185\pm0.046$ & a, d, a, a \\
$J0528+2200$ & $B0525+21$ & $3.746$ & $-13.4$ & $13.1$ & $31.5$ & $6.2$ & $1.215\pm0.304$ & $-0.115\pm0.029$ & a, a, e, e \\
$J0534+2200$ & $B0531+21$ & $0.033$ & $-12.4$ & $12.6$ & $38.6$ & $3.1$ & $2.000\pm0.500$ & $-0.168\pm0.042$ & a, n, m, a \\
$J0543+2329$ & $B0540+23$ & $0.246$ & $-13.8$ & $12.3$ & $34.6$ & $5.4$ & $1.565\pm0.391$ & $-0.059\pm0.015$ & a, a, c, a \\
$J0601-0527$ & $B0559-05$ & $0.396$ & $-14.9$ & $11.9$ & $32.9$ & $6.7$ & $2.083\pm0.521$ & $-0.453\pm0.113$ & a, d, b, d \\
$J0614+2229$ & $B0611+22$ & $0.335$ & $-13.2$ & $12.7$ & $34.8$ & $5.0$ & $3.571\pm0.893$ & $0.188\pm0.047$ & a, a, c, a \\
$J0629+2415$ & $B0626+24$ & $0.477$ & $-14.7$ & $12.0$ & $32.9$ & $6.6$ & $3.125\pm0.781$ & $0.376\pm0.094$ & a, a, c, a \\
$J0630-2834$ & $B0628-28$ & $1.244$ & $-14.1$ & $12.5$ & $32.2$ & $6.4$ & $0.320\pm0.080$ & $-0.065\pm0.016$ & a, a, b, a \\
$J0742-2822$ & $B0740-28$ & $0.167$ & $-13.8$ & $12.2$ & $35.2$ & $5.2$ & $2.000\pm0.500$ & $-0.055\pm0.014$ & a, a, e, a \\
$J0814+7429$ & $B0809+74$ & $1.292$ & $-15.8$ & $11.7$ & $30.5$ & $8.1$ & $0.432\pm0.108$ & $0.254\pm0.064$ & a, a, a, e \\
$J0820-1350$ & $B0818-13$ & $1.238$ & $-14.7$ & $12.2$ & $31.6$ & $7.0$ & $1.900\pm0.475$ & $0.445\pm0.111$ & a, a, b, a \\
$J0823+0159$ & $B0820+02$ & $0.865$ & $-16.0$ & $11.5$ & $30.8$ & $8.1$ & $2.500\pm0.625$ & $0.939\pm0.235$ & a, a, a, a \\
$J0826+2637$ & $B0823+26$ & $0.531$ & $-14.8$ & $12.0$ & $32.7$ & $6.7$ & $0.498\pm0.124$ & $0.290\pm0.073$ & a, a, c, e \\
$J0922+0638$ & $B0919+06$ & $0.431$ & $-13.9$ & $12.4$ & $33.8$ & $5.7$ & $1.100\pm0.275$ & $0.682\pm0.170$ & a, a, c, a \\
$J0953+0755$ & $B0950+08$ & $0.253$ & $-15.6$ & $11.4$ & $32.7$ & $7.2$ & $0.261\pm0.065$ & $0.208\pm0.052$ & a, a, a, a \\
$J1041-1942$ & $B1039-19$ & $1.386$ & $-15.0$ & $12.1$ & $31.1$ & $7.4$ & $2.533\pm0.633$ & $1.429\pm0.357$ & a, d, a, a \\
$J1136+1551$ & $B1133+16$ & $1.188$ & $-14.4$ & $12.3$ & $31.9$ & $6.7$ & $0.372\pm0.093$ & $0.375\pm0.094$ & a, a, c, e \\
$J1239+2453$ & $B1237+25$ & $1.382$ & $-15.0$ & $12.1$ & $31.2$ & $7.4$ & $0.840\pm0.210$ & $0.866\pm0.216$ & a, a, e, a \\
$J1509+5531$ & $B1508+55$ & $0.740$ & $-14.3$ & $12.3$ & $32.7$ & $6.4$ & $2.100\pm0.525$ & $1.688\pm0.422$ & a, a, a, a \\
$J1543-0620$ & $B1540-06$ & $0.709$ & $-15.1$ & $11.9$ & $32.0$ & $7.1$ & $3.226\pm0.806$ & $1.942\pm0.486$ & a, a, a, g \\
$J1607-0032$ & $B1604-00$ & $0.422$ & $-15.5$ & $11.6$ & $32.2$ & $7.3$ & $1.087\pm0.272$ & $0.655\pm0.164$ & a, a, c, a \\
$J1645-0317$ & $B1642-03$ & $0.388$ & $-14.7$ & $11.9$ & $33.1$ & $6.5$ & $3.846\pm0.962$ & $1.706\pm0.426$ & a, a, a, a \\
$J1705-1906$ & $B1702-19$ & $0.299$ & $-14.4$ & $12.1$ & $33.8$ & $6.1$ & $0.747\pm0.187$ & $0.193\pm0.048$ & a, a, a, a \\
$J1709-1640$ & $B1706-16$ & $0.653$ & $-14.2$ & $12.3$ & $33.0$ & $6.2$ & $0.562\pm0.141$ & $0.158\pm0.039$ & a, a, b, a \\
$J1713+0747$ & -- & $0.005$ & $-20.1$ & $8.3$ & $33.5$ & $9.9$ & $1.311\pm0.328$ & $0.582\pm0.146$ & p, f, f, h \\
$J1722-3207$ & $B1718-32$ & $0.477$ & $-15.2$ & $11.7$ & $32.4$ & $7.1$ & $2.928\pm0.732$ & $0.147\pm0.037$ & a, a, b, g \\
$J1730-2304$ & -- & $0.008$ & $-19.7$ & $8.6$ & $33.2$ & $9.8$ & $0.650\pm0.163$ & $0.093\pm0.023$ & j, f, f, j \\
$J1735-0724$ & $B1732-07$ & $0.419$ & $-14.9$ & $11.9$ & $32.8$ & $6.7$ & $6.667\pm1.667$ & $1.539\pm0.385$ & a, a, a, a \\
$J1740+1311$ & $B1737+13$ & $0.803$ & $-14.8$ & $12.0$ & $32.0$ & $6.9$ & $4.176\pm1.044$ & $1.560\pm0.390$ & a, a, c, a \\
$J1740-3015$ & $B1737-30$ & $0.607$ & $-12.3$ & $13.2$ & $34.9$ & $4.3$ & $0.400\pm0.100$ & $0.027\pm0.007$ & a, a, b, g \\
$J1745-3040$ & $B1742-30$ & $0.367$ & $-14.0$ & $12.3$ & $33.9$ & $5.7$ & $0.200\pm0.050$ & $0.023\pm0.006$ & a, a, b, a \\
$J1752-2806$ & $B1749-28$ & $0.563$ & $-14.1$ & $12.3$ & $33.3$ & $6.0$ & $0.200\pm0.050$ & $0.023\pm0.006$ & a, a, b, a \\
$J1759-2205$ & $B1756-22$ & $0.461$ & $-14.0$ & $12.4$ & $33.6$ & $5.8$ & $3.256\pm0.814$ & $0.063\pm0.016$ & a, a, a, a \\
$J1803-2137$ & $B1800-21$ & $0.134$ & $-12.9$ & $12.6$ & $36.3$ & $4.2$ & $4.400\pm1.100$ & $0.024\pm0.006$ & a, a, a, g \\
$J1807-0847$ & $B1804-08$ & $0.164$ & $-16.5$ & $10.8$ & $32.4$ & $8.0$ & $1.500\pm0.375$ & $0.168\pm0.042$ & a, a, b, a \\
$J1820-0427$ & $B1818-04$ & $0.598$ & $-14.2$ & $12.3$ & $33.1$ & $6.2$ & $2.857\pm0.714$ & $0.254\pm0.064$ & a, a, b, a \\
$J1822-2256$ & $B1819-22$ & $1.874$ & $-14.9$ & $12.2$ & $30.9$ & $7.3$ & $3.260\pm0.815$ & $-0.232\pm0.058$ & a, a, b, g \\
$J1823-3106$ & $B1820-31$ & $0.284$ & $-14.5$ & $12.0$ & $33.7$ & $6.2$ & $1.586\pm0.397$ & $-0.206\pm0.052$ & a, a, b, g \\
$J1825-0935$ & $B1822-09$ & $0.769$ & $-13.3$ & $12.8$ & $33.7$ & $5.4$ & $0.300\pm0.075$ & $0.033\pm0.008$ & a, a, o, a \\
$J1832-0827$ & $B1829-08$ & $0.647$ & $-13.2$ & $12.8$ & $34.0$ & $5.2$ & $5.200\pm1.300$ & $0.039\pm0.010$ & a, a, b, a \\
$J1833-0338$ & $B1831-03$ & $0.687$ & $-13.4$ & $12.7$ & $33.7$ & $5.4$ & $2.500\pm0.625$ & $0.119\pm0.030$ & a, a, a, a \\
$J1847-0402$ & $B1844-04$ & $0.598$ & $-13.3$ & $12.8$ & $34.0$ & $5.3$ & $3.419\pm0.855$ & $-0.039\pm0.010$ & a, a, b, a \\
$J1849-0636$ & $B1846-06$ & $1.451$ & $-13.3$ & $12.9$ & $32.8$ & $5.7$ & $3.849\pm0.962$ & $-0.152\pm0.038$ & a, a, a, a \\
$J1857+0943$ & $B1855+09$ & $0.005$ & $-19.7$ & $8.5$ & $33.7$ & $9.7$ & $1.200\pm0.300$ & $0.088\pm0.022$ & a, f, f, a \\
$J1900-2600$ & $B1857-26$ & $0.612$ & $-15.7$ & $11.6$ & $31.5$ & $7.7$ & $0.700\pm0.175$ & $-0.138\pm0.035$ & a, d, b, l \\
$J1901+0331$ & $B1859+03$ & $0.655$ & $-14.1$ & $12.3$ & $33.0$ & $6.1$ & $7.000\pm1.750$ & $-0.069\pm0.017$ & a, d, d, a \\
$J1903+0135$ & $B1900+01$ & $0.729$ & $-14.4$ & $12.2$ & $32.6$ & $6.5$ & $3.300\pm0.825$ & $-0.094\pm0.024$ & a, a, c, a \\
$J1909+1102$ & $B1907+10$ & $0.284$ & $-14.6$ & $11.9$ & $33.7$ & $6.2$ & $4.800\pm1.200$ & $0.099\pm0.025$ & a, a, a, a \\
$J1913-0440$ & $B1911-04$ & $0.826$ & $-14.4$ & $12.3$ & $32.5$ & $6.5$ & $4.041\pm1.010$ & $-0.485\pm0.121$ & a, a, b, a \\
$J1917+1353$ & $B1915+13$ & $0.195$ & $-14.1$ & $12.1$ & $34.6$ & $5.6$ & $5.882\pm1.471$ & $0.078\pm0.020$ & a, a, c, a \\
$J1921+2153$ & $B1919+21$ & $1.337$ & $-14.9$ & $12.1$ & $31.3$ & $7.2$ & $0.300\pm0.075$ & $0.045\pm0.011$ & a, a, c, a \\
$J1932+1059$ & $B1929+10$ & $0.227$ & $-14.9$ & $11.7$ & $33.6$ & $6.5$ & $0.310\pm0.077$ & $0.005\pm0.001$ & i, a, c, e \\
$J1932+2220$ & $B1930+22$ & $0.144$ & $-13.2$ & $12.5$ & $35.9$ & $4.6$ & $10.900\pm2.725$ & $0.303\pm0.076$ & a, a, a, a \\
$J1935+1616$ & $B1933+16$ & $0.359$ & $-14.2$ & $12.2$ & $33.7$ & $6.0$ & $3.700\pm0.925$ & $-0.115\pm0.029$ & a, a, c, a \\
$J1939+2134$ & $B1937+21$ & $0.002$ & $-19.0$ & $8.6$ & $36.0$ & $8.4$ & $3.500\pm0.875$ & $0.003\pm0.001$ & i, f, i, h \\
$J1941-2602$ & $B1937-26$ & $0.403$ & $-15.0$ & $11.8$ & $32.8$ & $6.8$ & $3.564\pm0.891$ & $-1.308\pm0.327$ & a, a, b, g \\
$J1946+1805$ & $B1944+17$ & $0.441$ & $-16.6$ & $11.0$ & $31.0$ & $8.5$ & $0.300\pm0.075$ & $0.008\pm0.002$ & a, d, a, d \\
$J1948+3540$ & $B1946+35$ & $0.717$ & $-14.2$ & $12.4$ & $32.9$ & $6.2$ & $7.650\pm1.913$ & $0.692\pm0.173$ & a, a, a, a \\
$J1954+2923$ & $B1952+29$ & $0.427$ & $-17.8$ & $10.4$ & $29.9$ & $9.6$ & $0.536\pm0.134$ & $0.034\pm0.008$ & a, d, a, a \\
$J1955+5059$ & $B1953+50$ & $0.519$ & $-14.9$ & $11.9$ & $32.6$ & $6.8$ & $2.186\pm0.546$ & $0.464\pm0.116$ & a, a, a, a \\
$J2002+4050$ & $B2000+40$ & $0.905$ & $-14.8$ & $12.1$ & $32.0$ & $6.9$ & $6.391\pm1.598$ & $0.611\pm0.153$ & a, a, a, a \\
$J2004+3137$ & $B2002+31$ & $2.111$ & $-13.1$ & $13.1$ & $32.5$ & $5.7$ & $8.000\pm2.000$ & $0.021\pm0.005$ & a, a, c, a \\
$J2013+3845$ & $B2011+38$ & $0.230$ & $-14.1$ & $12.2$ & $34.5$ & $5.6$ & $7.123\pm1.781$ & $0.329\pm0.082$ & a, a, a, a \\
$J2018+2839$ & $B2016+28$ & $0.558$ & $-15.8$ & $11.5$ & $31.5$ & $7.8$ & $0.980\pm0.245$ & $-0.042\pm0.011$ & a, a, c, a \\
$J2022+2854$ & $B2020+28$ & $0.343$ & $-14.7$ & $11.9$ & $33.3$ & $6.5$ & $2.100\pm0.525$ & $-0.146\pm0.037$ & a, a, c, a \\
$J2022+5154$ & $B2021+51$ & $0.529$ & $-14.5$ & $12.1$ & $32.9$ & $6.4$ & $1.800\pm0.450$ & $0.289\pm0.072$ & a, a, e, e \\
$J2048-1616$ & $B2045-16$ & $1.962$ & $-14.0$ & $12.7$ & $31.8$ & $6.5$ & $0.950\pm0.237$ & $-0.494\pm0.123$ & a, a, b, a \\
$J2055+3630$ & $B2053+36$ & $0.222$ & $-15.4$ & $11.5$ & $33.1$ & $7.0$ & $5.000\pm1.250$ & $-0.463\pm0.116$ & a, a, a, a \\
$J2113+2754$ & $B2110+27$ & $1.203$ & $-14.6$ & $12.3$ & $31.8$ & $6.9$ & $1.429\pm0.357$ & $-0.320\pm0.080$ & a, a, c, a \\
$J2113+4644$ & $B2111+46$ & $1.015$ & $-15.1$ & $11.9$ & $31.4$ & $7.4$ & $2.174\pm0.543$ & $-0.021\pm0.005$ & a, d, d, d \\
$J2145-0750$ & -- & $0.016$ & $-19.5$ & $8.8$ & $32.5$ & $9.9$ & $0.714\pm0.179$ & $-0.453\pm0.113$ & j, f, f, h \\
$J2157+4017$ & $B2154+40$ & $1.525$ & $-14.5$ & $12.4$ & $31.6$ & $6.8$ & $2.900\pm0.725$ & $-0.543\pm0.136$ & a, a, a, a \\
$J2219+4754$ & $B2217+47$ & $0.538$ & $-14.6$ & $12.1$ & $32.8$ & $6.5$ & $2.387\pm0.597$ & $-0.287\pm0.072$ & a, a, a, a \\
$J2225+6535$ & $B2224+65$ & $0.683$ & $-14.0$ & $12.4$ & $33.1$ & $6.0$ & $0.900\pm0.225$ & $0.135\pm0.034$ & a, d, a, a \\
$J2257+5909$ & $B2255+58$ & $0.368$ & $-14.2$ & $12.2$ & $33.7$ & $6.0$ & $3.000\pm0.750$ & $0.000\pm0.000$ & a, a, a, a \\
$J2305+3100$ & $B2303+30$ & $1.576$ & $-14.5$ & $12.3$ & $31.5$ & $6.9$ & $4.348\pm1.087$ & $-1.922\pm0.480$ & a, a, c, a \\
$J2313+4253$ & $B2310+42$ & $0.349$ & $-15.9$ & $11.3$ & $32.0$ & $7.7$ & $1.060\pm0.265$ & $-0.272\pm0.068$ & a, a, a, a \\
$J2321+6024$ & $B2319+60$ & $2.256$ & $-14.2$ & $12.6$ & $31.4$ & $6.7$ & $2.700\pm0.675$ & $0.004\pm0.001$ & a, d, e, a \\
$J2326+6113$ & $B2324+60$ & $0.234$ & $-15.5$ & $11.5$ & $33.0$ & $7.0$ & $2.733\pm0.683$ & $0.031\pm0.008$ & a, a, a, a \\
$J2354+6155$ & $B2351+61$ & $0.945$ & $-13.8$ & $12.6$ & $32.9$ & $6.0$ & $2.439\pm0.610$ & $0.022\pm0.006$ & a, a, a, a \\
\hline \hline 
\multicolumn{10}{p{.85\textwidth}}{\small Columns:  $P$: period; $\dot{P}$: period derivative; $B$: magnetic field strength; $\dot{E}$: spin-down energy; $\tau_c$: derived age; Dist.: best distance estimate based on the YMW16 electron-density model~\citep{Yao2017ApJ...835...29Y}; Dist.${\rm GP}$: distance from Galactic plane based on YMW16; References for pulsars in FB1, FB2, FB3, and FB4, respectively. Values taken from the ATNF pulsar catalogue~\citep{Manchester+2005AJ....129.1993M}. a: \citet{gl98}, b: \citet{jk17}, c: \citet{Weisberg1999ApJS..121..171W}, d: \citet{antt94}, e: \citet{hx97b}, f: \citet{dhm+15}, g: \citet{wmlq93}, h: \citet{kxl+98}, i: \citet{stc99}, j: \citet{lor94}, k: \citet{bbm+97}, l: \citet{mhq98}, m: \citet{mh99}, n: \citet{lun94}, o: \citet{kj06}, p: \citet{fwc93}.}
\end{longtable}
\onecolumn
    
\newpage
\section{MST variants}
\label{app:other-msts}

\begin{figure*}[h]
\includegraphics[trim=0 0 0 0, clip, width=17cm]{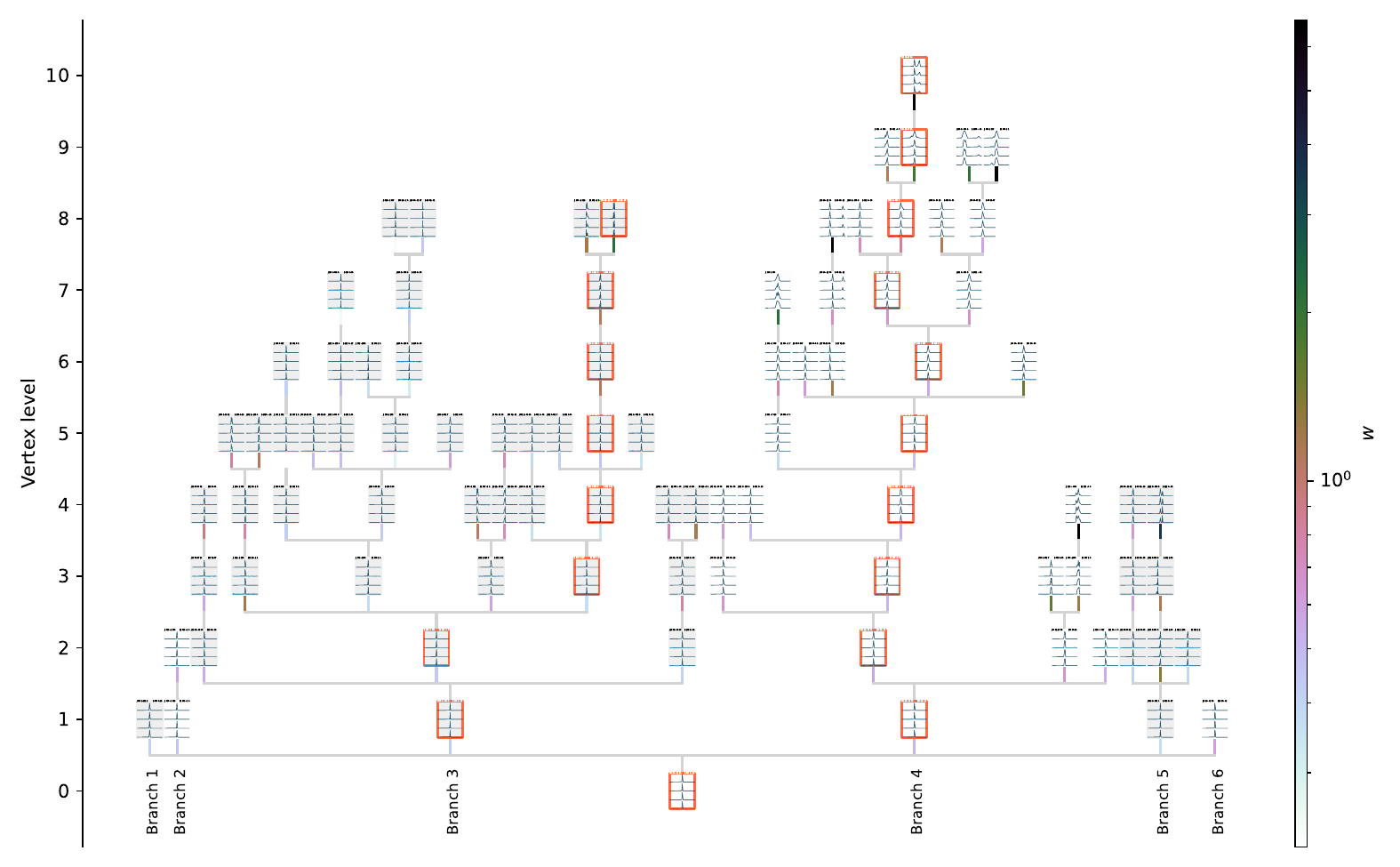}
\caption{Minimum spanning tree from Figure \ref{fig:tree-4freqs-w10} (lower panel) but shown here with full phase information, $\phi\in[-0.5, 0.5]$, where cases with an interpulse can now be seen.}
\label{fig:tree-full}%
\end{figure*}

\begin{figure*}[h]
\includegraphics[trim=0 0 0 0, clip, width=17cm]{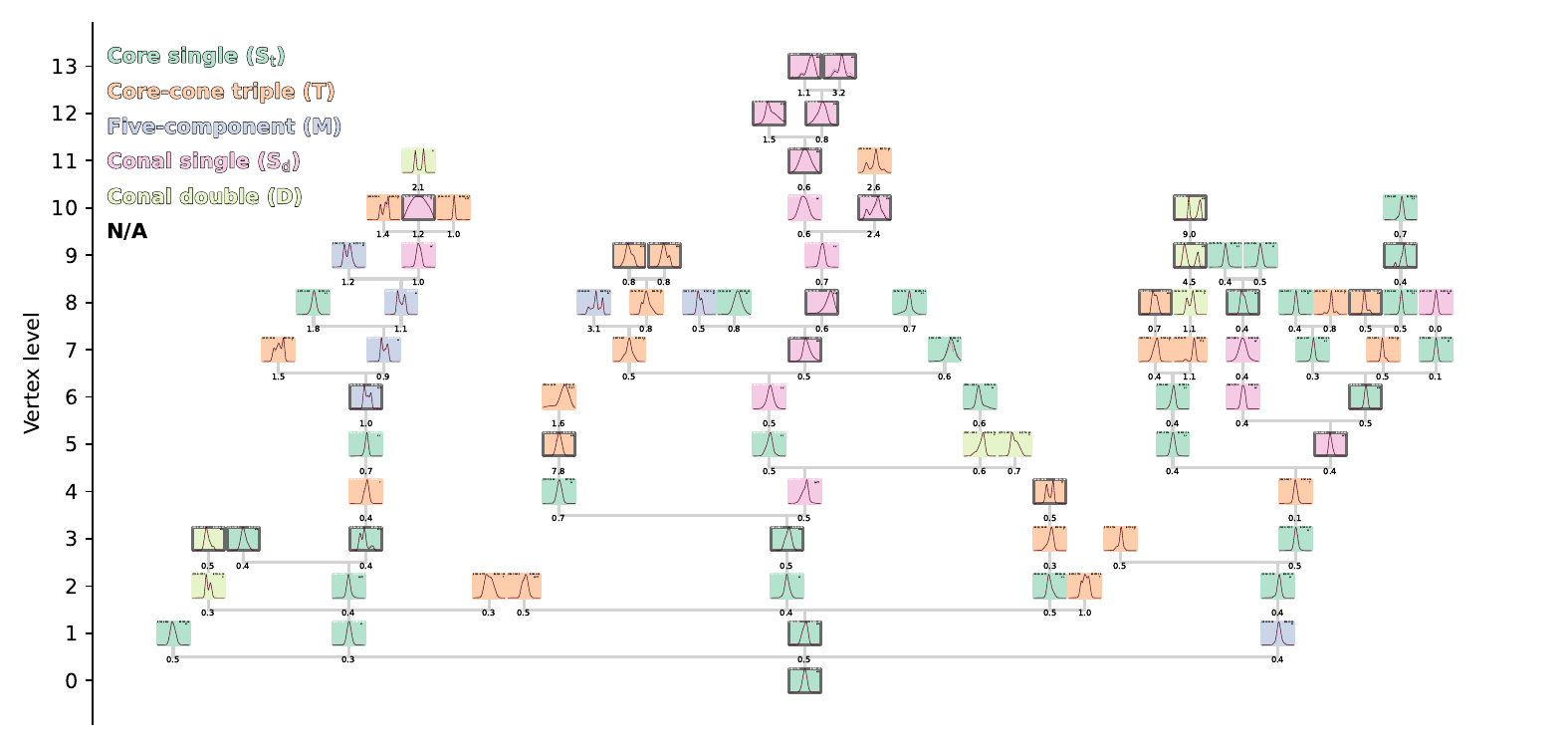}
\caption{Classification using profiles in the 400--700\,MHz bin.}
\label{fig:classification-FB1}%
\end{figure*}

\begin{figure*}[h]
\includegraphics[trim=0 0 0 0, clip, width=17cm]{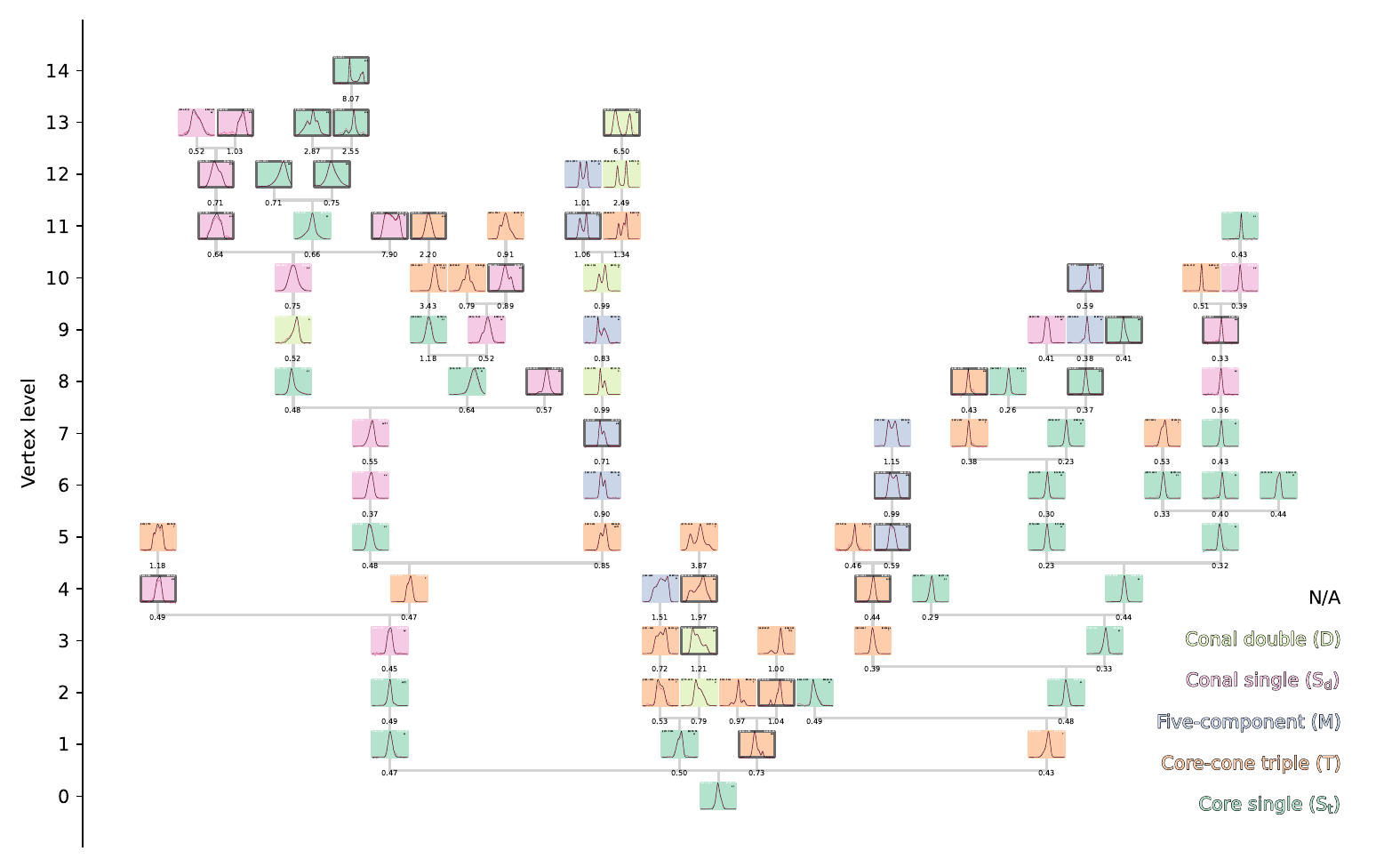}
\caption{Classification using profiles in the 700--1000\,MHz bin.}
\label{fig:classification-FB2}%
\end{figure*}

\begin{figure*}[h]
\includegraphics[trim=0 0 0 0, clip, width=17cm]{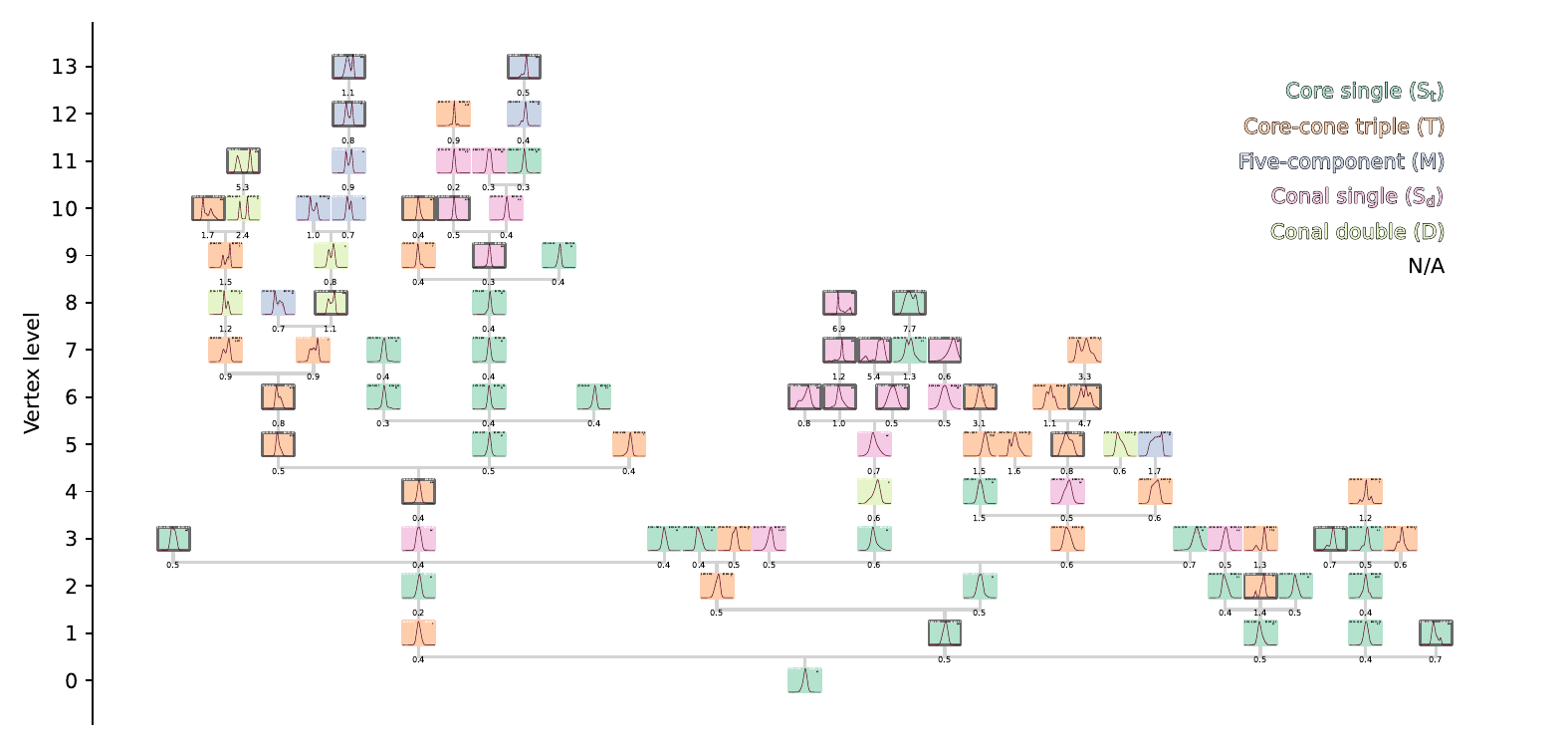}
\caption{Classification using profiles in the 1000--1500\,MHz bin.}
\label{fig:classification-FB3}%
\end{figure*}

\begin{figure*}[h]
\includegraphics[trim=0 0 0 0, clip, width=17cm]{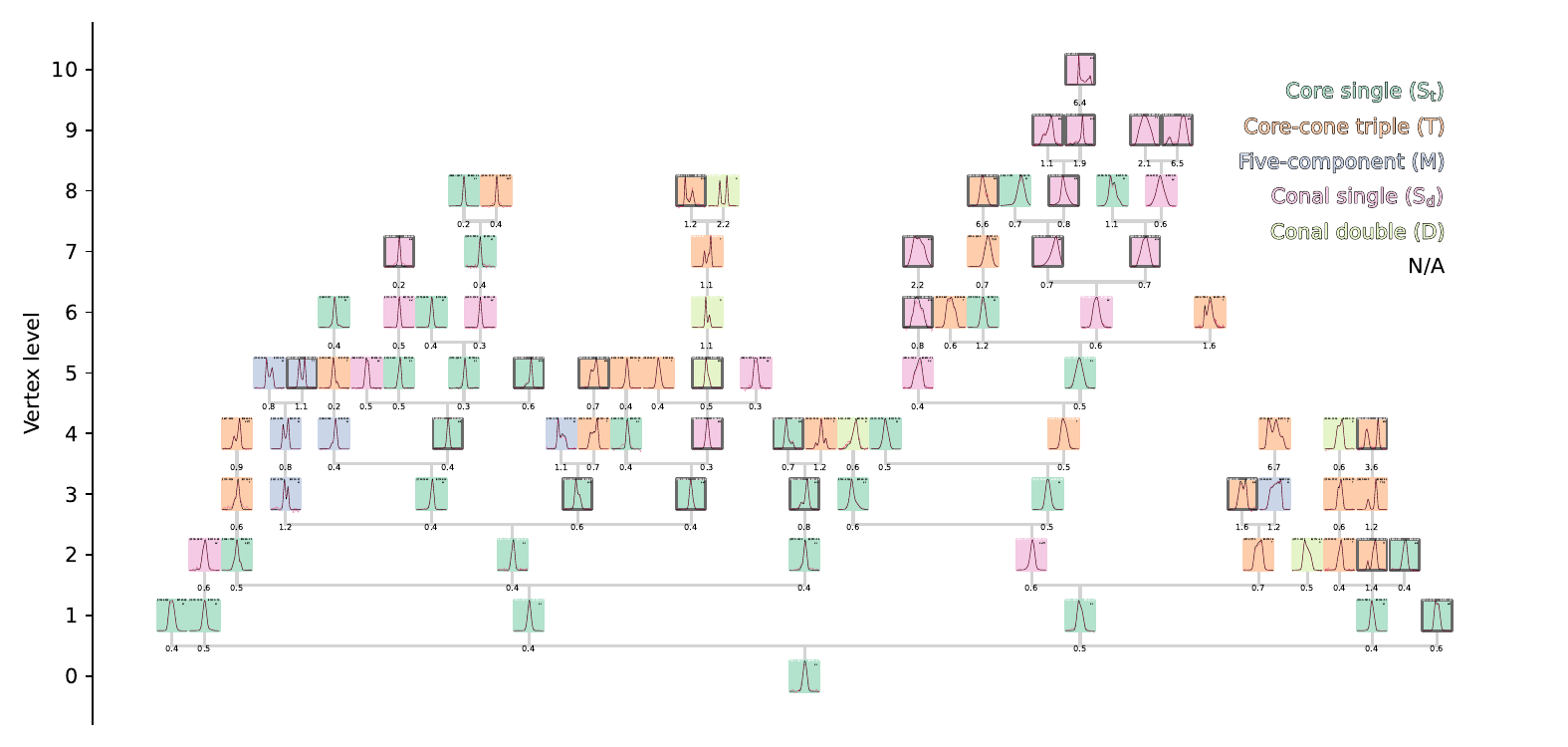}
\caption{Classification using profiles in the 1500--2000\,MHz bin.}
\label{fig:classification-FB4}%
\end{figure*}

This section presents MST variants discussed in Sections \ref{sec:results} and \ref{sec:classification}. 
In particular, Figure \ref{fig:tree-full} presents Figure \ref{fig:tree-4freqs-w10} using the full phase information. 
Figures \ref{fig:classification-FB1} to \ref{fig:classification-FB4} present MSTs obtained using individual frequency bins and their corresponding classification results listed in Table \ref{tab:classes}.

\section{Spin parameters of pulsar subsets}

In this section, we investigate the difference between the two main branches of the MST presented in Figure \ref{fig:tree-4freqs-w10}, namely branches 3 and 4.

\begin{figure*}[h]
\includegraphics[trim=0 0 0 0, clip, width=\columnwidth]{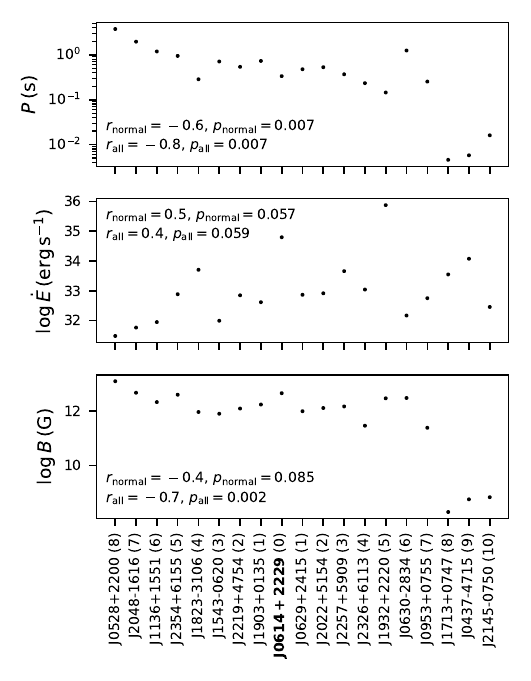}
\caption{Pulsar spin parameters ordered as a function of the longest sequence  (Fig. \ref{fig:sequence_multifreq}). Pulsar names (J2000.0) are accompanied by their MST vertex level in parentheses.}
\label{fig:sequence-versus-spinparams}%
\end{figure*}

Firstly, Figure \ref{fig:sequence-versus-spinparams} presents the evolution of spin parameters $P$, $\dot{E}$ and $B$ as a function of the pulsar order within the longest sequence discussed in Section \ref{subsec:multifreq_longest}. 
There is a general trend for pulsars period varying from slow pulsars in branch 3 (left to the root PSR J0614+2229) towards faster pulsars in branch 4.
Similarly, there appear to be a  correlation between $\dot{E}$ and sequence order, while $B$ appears near constant throughout, with the exclusion of the three millisecond pulsars at the right end.
To evaluate the monotonicity of the relationship between each physical parameter and the sequence index, we indicate the Spearman correlation coefficient $r$ and related $p$-value in each panel. 
We evaluate the correlation for two scenarios: normal pulsars ($P>0.1$\,ms) and for all pulsars in the sequence. 
Results show that these parameters are partially monotonically correlated. 

Secondly, columns (1) and (2) of Figure \ref{fig:sequence-versus-spinparams} present cumulative distributions for $P$, $\dot{P}$, $\dot{E}$, $B$, and Age, where column (1) includes all its pulsars, and (2) includes only pulsars with $P \ge 0.1$\,ms.  
We perform two sample Kolmogorov-Smirnov (KS) tests between the samples of both branches. 
While distributions in terms of $P$, $\dot{P}$, $B$, and age appear as significant when considering all pulsars, when $p$-values are approximately less than or equal to 0.01, only a marginal difference can be found in the spin period when considering normal pulsars only. 
Finally, column (3) compares two sub-branches of branch 3: one including primarily single-component profiles and the other having more complex profiles. 
Based on their respective cumulative distributions, these two sub-branches appear to be drawn from the same distribution.

\begin{figure*}[h!]
\centering
\begin{tabular}{ccc}
\includegraphics[trim=0 0 0 0, clip, width=5.7cm]{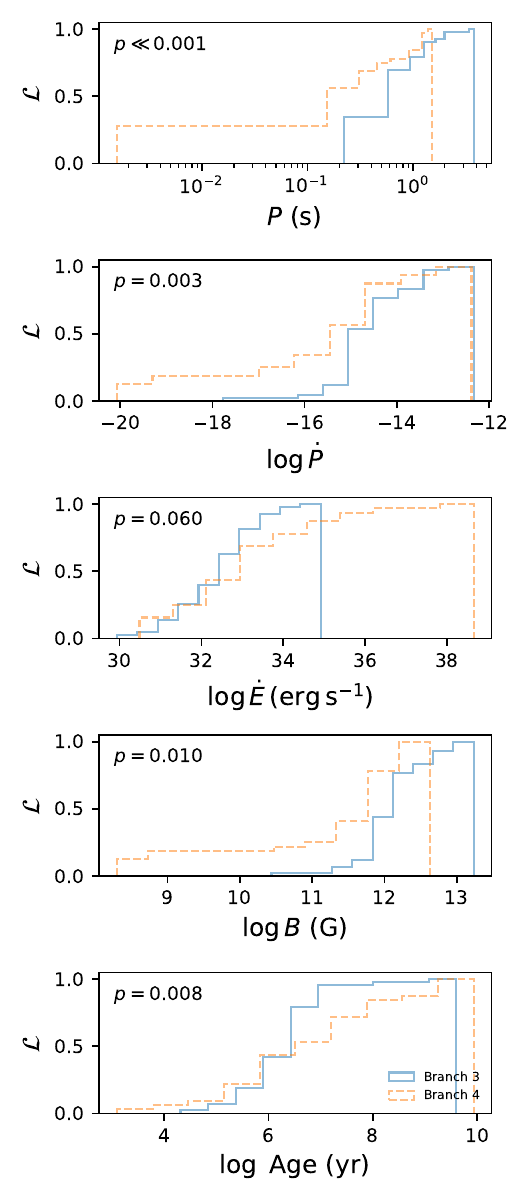} & 
\includegraphics[trim=0 0 0 0, clip, width=5.7cm]{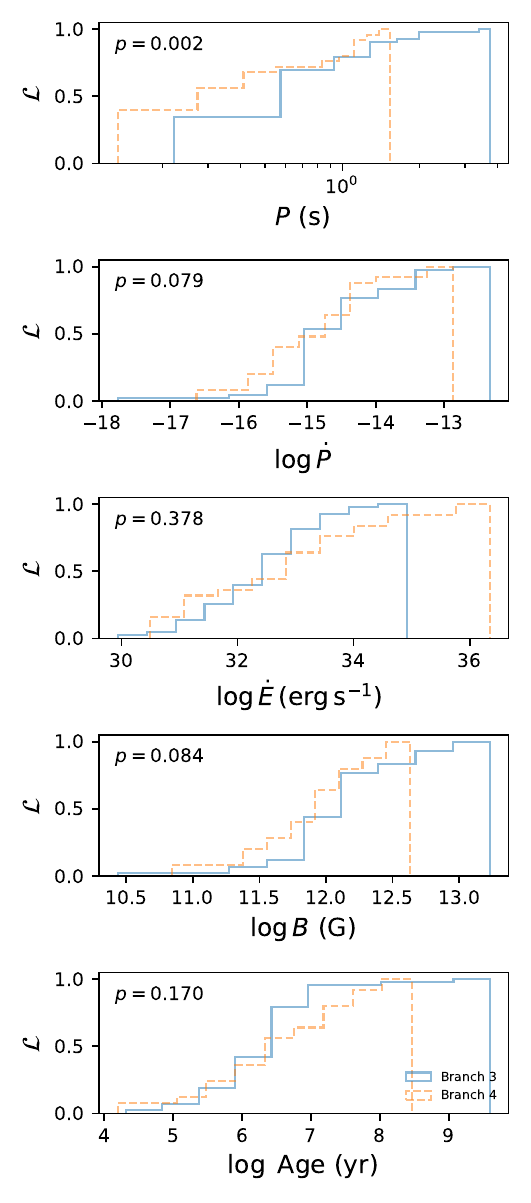} &
\includegraphics[trim=0 0 0 0, clip, width=5.7cm]{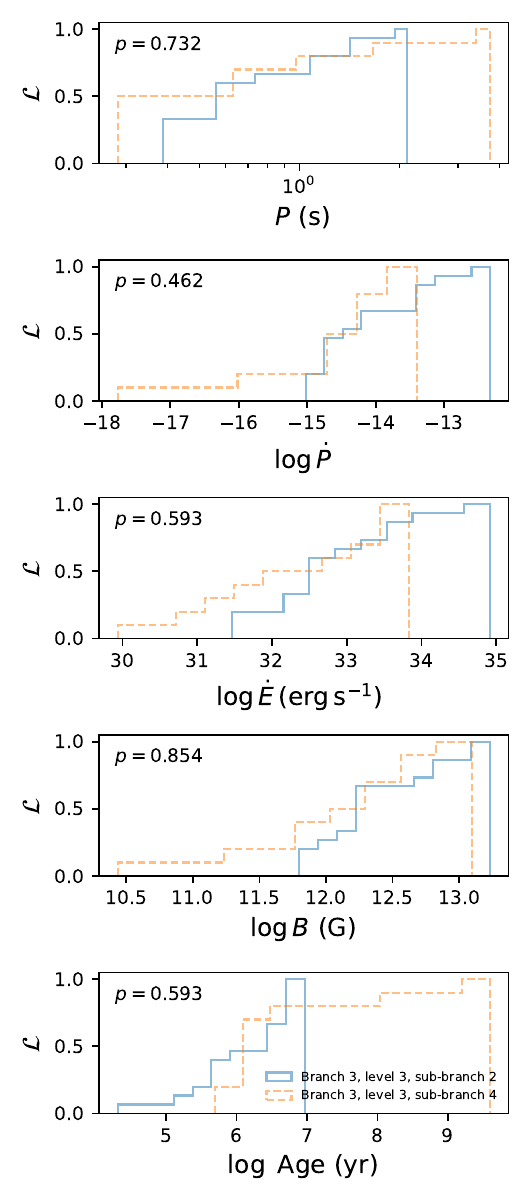} \\
(1) & (2) & (3) \\
\end{tabular}
\caption{Cumulative distributions of spin parameters for (1) branches 3 and 4; (2) branches 3 and 4 while excluding millisecond pulsars; and (3) branch 3, level 3, sub-branches 2 and 4. We indicate the $p$-values obtained by performing a two-sample Kolmogorov-Smirnov test between the samples of both branches compared in each panel. Marginal differences can be found on the spin period between branches 3 and 4 when considering normal pulsars only.}
\label{fig:cumulative_dists_branches}%

\end{figure*}

\section{Distance uncertainty}
\begin{figure*}[t]
\centering
\includegraphics[trim=0 0 0 0, clip, width=135mm]{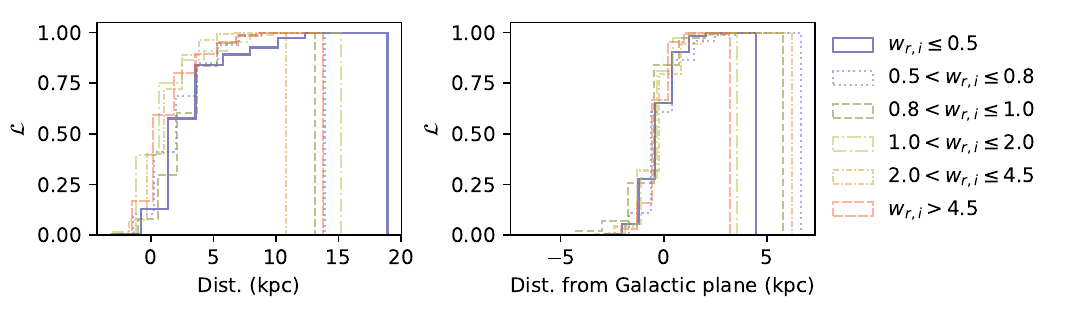}
\caption{Cumulative distributions for distances from 10,000 realisations per measured distance from our 90 pulsars sampled from a normal distribution around the distance uncertainty.}
\label{fig:cumulative-dists-w-uncertainties}%
\end{figure*}

As mentioned in Appendix \ref{app::pulsar_set}, distance to pulsars used in this work are taken from the ATNF pulsar catalogue. 
These distances are based on YMW16, and we represent distance uncertainty using a fiducial factor of 0.25.
Here, we want to visualise the effect of uncertainties on the cumulative distributions presented in Figure \ref{fig:cumulative-dists}. 
Given distances $d \in D$, we recompute the cumulative distribution after generating 10,000 realisations randomly sampled from a normal distribution of mean $d$, with standard deviation $d/4$. 
Results are shown in Figure \ref{fig:cumulative-dists-w-uncertainties}. 

\end{appendix}

\end{document}